# Only skin deep? Depth-resolved photo-oxidation of 3D printed polymers PLA, ABS, PETG, HIPS and PC after 322 days of continuous UV exposure

Jan Tomastik[a,b*], Karolina Siskova[c], Lukas Vaclavek[a,b], Radim Ctvrtlik[a]

[a]*Palacký University in Olomouc, Faculty of Science, Joint Laboratory of Optics of Palacký University and Institute of Physics AS CR, 17. listopadu 12, 779 00, Olomouc, Czech Republic*

[b]*Institute of Physics of the Czech Academy of Sciences, Joint Laboratory of Optics of Palacký University and Institute of Physics of the Czech Academy of Sciences, 779 00, Olomouc, Czech Republic*

[c]*Department of Experimental Physics, Faculty of Science, Palacký University Olomouc*

**tomastik@jointlab.upol.cz*

**Abstract:**
This study investigates the effect of long-term UV radiation exposure on the mechanical and structural properties of selected polymeric materials – PLA, ABS, ABST, PETG, HIPS and PC – fabricated using 3D-print based Fused Filament Fabrication (FFF) method. Existing research in the field of polymers weathering has predominantly examined combined degradation factors (UV irradiation, humidity and thermal cycling) over the timescales rarely exceeding a few months, which prevents attribution of property changes to individual agents. To isolate the contribution of UV radiation specifically, samples were exposed up to 7744 consecutive hours (i.e. 322 days) of UV irradiation in an environmental cell with controlled humidity and temperature. The mechanical properties were characterized using nanoindentation, which allows for spatially resolved measurements of hardness and elastic modulus in the near-surface region, thereby enabling the detection of degradation gradients on the order of tens of micrometers. The lack of methodological consistency in the current literature hinders the cross-material comparison; this study addresses the gap by applying nanoindentation uniformly across wide range of selected materials prepared and tested under identical conditions, while studying these properties after a very long exposure to UV radiation.

Based on our analysis of mechanical properties, PLA showed no measurable change in hardness or reduced modulus over the full exposure period. PC showed a smaller increase than the other materials, which could not be clearly separated from the drift measured on stored control samples. ABS and HIPS underwent surface hardening with a mechanical properties' gradient extending beyond 10 µm, while for PETG, the most strongly affected material, the gradient was confined to a near-surface zone shallower than 10 µm. The increases in hardness and modulus were accompanied by the development of surface crack networks observed by confocal microscopy, indicating a change in fracture behaviour. For PLA, ABS and PETG, vibrational spectroscopies (Raman scattering and infrared absorption) were used to link mechanical property observations to structural changes at the molecular level – PLA showed no spectral changes upon UV exposure, while ABS and PETG exhibited modifications in both characteristic and deformation vibration bands, consistent with photooxidative degradation of their respective polymer chains.

**Keywords:** polymer degradation; UV degradation; 3D printed polymer; photo-induced degradation; mechanical properties; structural properties; nanoindentation; indentation creep

**Highlights:**

- Separating UV exposure from thermal and chemical agents is a prerequisite for targeted improvement of polymer resistance; here UV was isolated as the single variable
- Long-term UV-induced degradation of FFF 3D-printed polymers remains poorly characterised at the surface scale
- Nanoindentation sensitively detects photo-oxidative surface changes in polymers over a depth range from nanometres to tens of micrometres
- PLA retains its mechanical properties without measurable change after 7744 hours (322 days) of continuous UV irradiation
- PETG and ABS-T show the strongest UV-induced surface mechanical changes, with the property gradient confined to a near-surface zone shallower than 10 µm
- Surface hardening in ABS, PC and HIPS, accompanied by the formation of surface cracks, progresses only slightly between 3872 and 7744 h of UV exposure

## 1 Introduction

Significant step in polymer usage and application, besides its industrial production, has occurred in last decade as laboratories and general population has been introduced to the 3D printing technology [1]. Nowadays, 3D printing is employed in every scientific department for small tools production and even in many households due to increasingly accessible printers. The importance of independent production of plastic parts and/or components in companies and/or hospitals was demonstrated especially in the period of Covid-19 pandemic [2, 3].

Mechanical and structural properties of 3D printed plastic products are determined primarily by the choice of the printing material. There is a number of recommendations for print settings for each type of polymer used [4-9]. Although plastics have a low strength compared to the values of metals, they exhibit a higher strength per unit weight [10, 11]. The only problem connected with polymers is their lifespan that has begun to be an environmental issue during last several decades. Current methods of recycling or composting are still insufficient given the absolute volumes of plastic used globally, and the processing of plastic waste is thus an ever-increasing problem [12-14]. Therefore, it is essential to avoid new plastic waste generation. One route is the functional life extension of plastics [15-17], which requires an understanding of how they degrade in service.

With extensive spreading of 3D printed plastic products and current overuse of polymers, a proper understanding of their ageing mechanisms is needed for both, recycling and prolonging their functionality. In general, ageing or weathering mechanisms of polymers can be divided into physical and chemical. The former is manifested through densification due to a molecular relaxation caused by non-equilibrium cooling during production of materials, or by processes leading to surface embrittlement. The latter is caused by oxidation, adsorption of pollutants, or chemical reactions between material components or with the reactive environment leading to structural changes [18]. In general, degradation processes are caused or accelerated by the weathering, which occurs due to exposure to temperature changes, humidity, or UV component of solar radiation. Research on the weathering of polymers and related changes in mechanical properties has been carried out on several materials in the past, but it was mostly a combination of several effects - UV exposure, temperature cycling and exposure to moisture [19, 20]. Such a study is motivated by the effort to control materials stability which can be either increased (by specific processes), or, conversely, decreased with an effort to biodegrade some types of plastics [21, 22]. Furthermore, for the sake of a direct comparison of materials´ mechanical properties, it is tremendously important to use the same type of mechanical tests and approaches.

Organic polymers exposed to sunlight are subjected to photochemical and photophysical degradation, caused mainly by the near-UV component with wavelengths between 290 and 400 nm [23]. The absorbed energy of the UV radiation causes excitation in the electronic structure of the polymer molecule. This can lead to re-emission of energy through light or heat or to changes in macro-molecular bonds [24]. Incident radiation induces deformation of the atomic bond in the chain of the polymer macromolecule [25]. Radiation of wavelength below 350 nm has sufficient energy to cause photo-scission of macromolecules to smaller pieces, which tend to react with oxygen, creating free radicals. Specific wavelength with maximum damaging effect is slightly different for various materials due to the bonds present in their structure, for example 300 nm for polyethylene (PE) and 370 nm for polypropylene (PP)[16]. Oxygenated radicals initiate further degradation of the polymer structure through chain-scission or an opposite phenomenon of crosslinking [26]. These reactions begin at the surface of the material, but with longer exposure to UV radiation, the material degrades more in depth. This results in vast changes in the structural and mechanical properties of the polymers, like reduction in hardness and modulus of elasticity, increase of brittleness leading to decrease of impact resistance, etc. [27].

The aim of this study is to isolate the contribution of UV radiation to the degradation of surface mechanical and structural properties of selected FFF 3D-printed polymers. This is deliberately a single-factor experiment and not a weathering simulation: specimens were exposed in a light-tight environmental cell in which UV irradiation was the only variable or weathering agent, while temperature, relative humidity and atmosphere were held constant, with no thermal cycling, condensation or ambient daylight. Correlation with outdoor service life is therefore not claimed and lies outside the scope of this work. What the experiment does provide is an unambiguous attribution of the observed changes to photooxidation, over an exposure substantially longer than in previously reported studies.

While UV radiation-induced photooxidation of polymers is primarily a surface phenomenon, it is desirable to study its effect deeper below the surface layer. This can be performed through analysis of structural and mechanical properties and their gradient from the surface to the depth of the sample. For this reason, a locally-sensitive instrumented depth-sensing nanoindentation method was used in this study. Nanoindentation is a precise instrumentally based method developed for evaluation of local mechanical properties, particularly the hardness and modulus of elasticity. It is commonly used for measurement of thin films and coatings [28], micro-objects [29] and bulk materials at room, or even at elevated temperatures [30].

Structural changes induced by photo-oxidation in polymers are tightly connected with changes in mechanical properties. Despite being a microscopic phenomenon, molecular scission results in increased brittleness, usually accompanied by an increase in hardness, resulting in the formation of microscopic cracks that can lead to surface inhomogeneity or eventual product breakage. Therefore, the study of mechanical properties using nanoindentation was supported by the evaluation of structural changes using vibrational spectroscopic techniques, namely Raman scattering (RS) and Infrared (IR) absorption. Both techniques can provide characteristic spectra that are unique to species at the molecular level. While IR absorption is often used in investigation of polymer structural changes induced by weathering [31-38], RS has been exploited less frequently [39-41]. The combination of IR and RS has been employed only in recent studies [42-47]. Hence, RS and IR absorption were both used in this study to evaluate the changes in the selected polymers after long-term UV exposure.

It should be stressed that despite the enormous application potential and extensive effort, there is still a lack of information about the real long-term UV exposure. There are plenty of studies on degradation [16, 24-26, 32], but they mostly address only structural aspects without evaluation of mechanical properties. On the other hand, the surface deterioration and the formation of microcracks can lead to macroscopic failures of the printed object.

Using the above techniques, several polymeric materials were investigated within this study: polylactic acid (PLA), acrylonitrile-butadiene-styrene (ABS), polyethylene terephthalate with glycol additive (PETG), high impact polystyrene (HIPS), and polycarbonate (PC). The selected polymers were chosen based on their widespread use, history, and availability to the 3D print FFF community. Details about their history and reported photosensitivity can be found out in the literature [23, 33, 48-54].

Given the time scale of chosen UV exposure and the use of a comparative nanoindentation method, this study provides relevant and valuable data for a wide range of 3D printed polymeric samples.

# 2 Experimental section

## 2.1 Samples preparation

3D printed samples from different types of thermoplastic polymers were prepared using Fused filament fabrication method. The complete list of the investigated samples is shown in Table 1. For ABS and PLA samples, black and white colour variant were investigated (denoted as "w" and "b") to explore the possible difference in their interaction with UV radiation. Doped variants were also included in our study, namely ABS-T, which contains methyl methacrylate (MMA) as confirmed by the manufacturer on enquiry, although the concentration was not disclosed, and polyethylene terephthalate with glycol additive (PETG). To correlate mechanical properties of the same material from different companies, PLA and PETG samples from other manufacturers (marked as "2" and "3") were tested under the prescribed sample conditions.

Table 1. Description of 3D printed polymeric samples

| Sample identification | Sample description | Manufacturer |
|---|---|---|
| PLA w | polylactic acid; white | Plasty Mladeč |
| PLA w2 | polylactic acid; white (different manufacturer) | Plasty Mladeč |
| PLA b | polylactic acid; black | Plasty Mladeč |
| ABS w | acrylonitrile-butadiene-styrene; white | Plasty Mladeč |
| ABS b | acrylonitrile-butadiene-styrene; black | Plasty Mladeč |
| ABST b | acrylonitrile-butadiene-styrene with methyl methacrylate (concentration not disclosed); black | Prusa 3D print |
| PETG b | poly(ethylene terephthalate), glycol-modified; black | Prusa 3D print |
| PETG b2 | poly(ethylene terephthalate), glycol-modified; black (different manufacturer) | Plasty Mladec |
| PETG b3 | poly(ethylene terephthalate), glycol-modified; black (different manufacturer) | Plasty Mladec |
| HIPS w | high impact polystyrene; white | Plasty Mladeč |
| PC w | Polycarbonate; white | Plasty Mladeč |

Samples were produced directly at the workplaces of major manufacturers of materials for 3D printing in accordance with our required parameters using the printing setup (print speed, layer thickness) and temperatures (nozzle and bed) standard for each of the materials. For instance, ABS and PLA were printed with layer thickness 0.2 mm; bed/nozzle temperatures 90/245 °C for ABS and 55/215 °C for PLA. The samples were prepared as 1x1x1 $cm^3$ cubes. A 100 % fill was chosen to prevent voids causing loss of support and affecting the indentation test.

Before measurements. All samples were glued and marked in the matrix to an aluminium holder for easier handling.

## 2.2 UV-radiation exposure

Samples were glued to a holder for easier handling and placed in a custom-built environmental chamber equipped with a single 18 W T8 fluorescent UV lamp (NARVA LT 18W T8/UVS-2, supplied by UVC Servis) [55]. The samples were placed at the distance of 20 cm from the UV lamp, see Fig. 1. The initial indentation measurements were performed on unaffected samples, and further analyses were performed on same samples after 3872 hours (161 days) and finally after 7744 hours (323 days) of UV light exposure. The continuous UV exposure was interrupted only during analysis of the samples, after which the irradiation in the environmental cell continued according to the plan. It is worth noting that samples were also measured in the finer manner after 1, 4, 10, 34 and 200 hours of continual UV exposure; none of them showed differences compared to the unexposed state, and these points are therefore not discussed in the Results. The temperature of 23 °C and humidity of 45 – 55 % were kept constant.

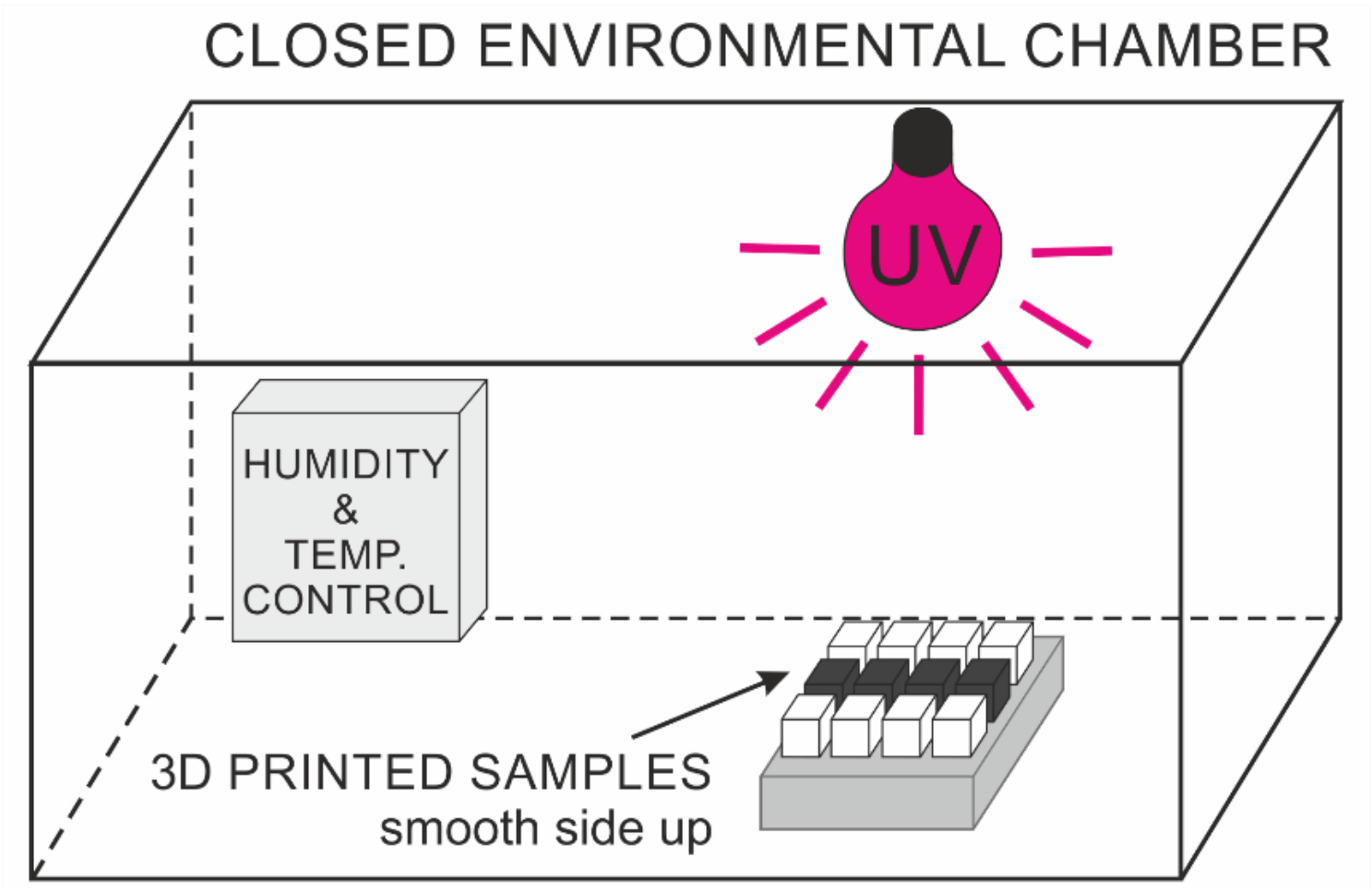

Fig. 1. Mounting of samples during UV exposure in an environmental chamber.

The experiment was designed to isolate photooxidation from the other agents acting simultaneously during outdoor weathering. No thermal cycling, condensation or water spray was applied, and the chamber was enclosed in opaque fabric so that the UV lamp was the only radiation source. Control specimens from the same printed batches were stored for the same period at 22 °C in a closed box inside a light-tight cabinet and re-measured, in order to separate physical ageing from photooxidative effects (see Section 3.4). The spectral and spatial characterisation of the radiation source was remeasured after the experiment and discussed in Section 3.2.

Visual (decolourization) and structural (surface homogeneity and microcracks) changes were then observed using laser scanning confocal microscopy (OLS LEXT 5000, Olympus) in various magnifications and several imaging modes.

## 2.3 Nanoindentation testing

Mechanical properties of samples before and after irradiation were measured using nanoindentation test on a NanoTest (MicroMaterials, UK, Ltd). The indentation tests require the lowest possible roughness of the examined surface, therefore, the tests were always performed on the bottom part of the cube which was in contact with the heated bed during printing, thus ensuring the smoothest possible surface. The standardized smoothing technique of chemical etching was not used as the resulting parasitic surface layer would have distorted the mechanical property measurements. The same surface was placed at the top on the sample holder inside the environmental chamber, facing the UV light in order to expose it to the radiation, see Fig. 1.

Four indentation loads of 0.5, 2, 10 and 100 mN were selected for each measured sample. This approach leads to different indentation depths and enables to monitor the gradient of changes in mechanical properties after exposure to UV radiation. In this way, indentation depths around 200 nm, 500 nm, 1000 – 1500 nm up to 4000 – 6000 nm were achieved for respective loads. 10 – 15 indentations at different locations were made for each load with 30 µm offset. Only standard shape curves were analysed; curve reduction ranged from 0 to 40 %. More significant reduction of non-standard shape curves was necessary for the shallowest imprints corresponding to the smallest indentation load of 0.5 mN, which were more affected by surface roughness and inhomogeneity. In such cases, defects on the surface have comparable dimension as indent itself which complicated their correct evaluation [56].

Complementary, H and E depth-profiling was performed via a load-partial unload indentation test (LPU). LPU experiments were performed using load range 5 – 200 mN, segmented into 30 cycles with "loading-holding-partial unloading" phases lasting 5 – 10 – 5 seconds. Experiments were repeated four times on each sample. Curves of non-standard shape were excluded from the analysis. All curves were then evaluated using standard nanoindentation procedure [57].

Very shallow indentations are more significantly affected by surface inhomogeneities, impurities and roughness than deeper ones. In our data, this led mainly to an increase in the standard deviation for low loads and to the occasional necessity of discarding indentation curves of non-standard shape from the evaluated set (with typically 10-20 curves 0-20 % at most). This effect is less pronounced for higher loads, which is reflected in a decrease in the standard deviation.

For each polymer one printed sample (one cube) was available firstly for an unexposed measurement, then UV exposed, and then remeasured in each step of UV exposure campaign – while 3872 hours and 7744 hours points are described thoroughly in this article (as they were the longest and most prominent ones). Another cube of each polymer was kept reserved for the control set. It is worth noting that measurement of Original (unexposed) samples was also performed on control set and while no differences (except expectable inhomogeneity) were observed, samples were thus regarded as homogenous and fully comparable. In general, the reported indentations represent repeat measurements at well-separated positions on a single specimen, and not independent manufacturing replicates. No formal statistical test was applied; the uncertainties quoted throughout are standard deviations of these repeats, and differences are discussed relative to the experimental scatter.

## 2.4 Creep measurement

We used nanoindentation tests with sufficiently long holding at maximum load to monitor creep of samples. Tests were conducted under indentation load of 20 mN with an extended holding time of 300 seconds during the maximum load stage. Ten measurements were always taken for each experimental setup; curves of non-standard shape were excluded from the analysis (in general 0 – 30 %). The slope of the curve was determined by tangent to the region of its last sixth, i.e. in the section between the 250th and 300th second of the creep measurement period.

Based on our detailed discussion about creep measurements provided in **Supporting Information** and its Fig. SI7, three creep parameters were reported for examined samples as follows:

(1) slope of the final phase of dh/dt curve
(2) relative increase in depth during creep
(3) parameter $P_{CR}$

While parameter (2) is an ISO standard [58] for polymeric samples from both tension and compression creep tests, it mainly describes range of primary creep phase. The parameter (1) can describe more accurately the latter part creep and predict its future trend. Goodal et al. [59] proposed the novel **parameter $P_{CR}$** (3) accounting the both aspects – the range of the creep for the specified test conditions (temperature, indentation load and length of the hold period) and the slope of the final phase of the observed creep. If the $P_{CR}$ is low than the creep in material was and should have remain low in the longer time span [60].

## 2.5 Structural examination using IR and Raman spectroscopy

Vibrational spectra recorded in this study were collected (a) on a FTIR spectrometer Nicolet iS5 from Thermo Scientific and (b) on a ProRaman L spectrometer TSI ChemLogix, equipped with diode laser of 785 nm wavelength (power output adjustable between 0 – 300 mW, electronic laser shutter control), high throughput fiber optics probe (O.D. > 8 at laser wavelength), and high sensitivity, ultra-low noise CCD spectrograph for 785 nm excitation (thermoelectrically cooled CCD detector to − 60 °C). The presented vibrational spectra (both IR and RS).

In case (a), attenuated total reflection (ATR) mode on a diamond crystal was used in the range of 400-4000 $cm^{-1}$ with the resolution of 4 $cm^{-1}$ and 64 scans. ATR mode is more convenient than classical arrangement (transmission mode) to study the surface of the samples due to the limited penetration depth of the evanescent wave [34]. Before each sample, background spectrum was recorded and automatically subtracted from the sample signal to eliminate signal of water and carbon dioxide stemming from ambient atmosphere (air). Omnic software was exploited for spectra acquisition and basic treatment which includes advanced ATR correction and automatic baseline correction. Any smoothing and/or further treatment of IR spectra were avoided. IR spectra are presented as the function of absorption on wavenumbers since it is more convenient for the quantification when carbonyl index is estimated.

In case (b), RamanReader®-L7B1 instrument control and data collection software was exploited; each RS spectrum was collected in a 180° orientation with 1s exposition time and 120 repetitions. The laser power was set to the maximum (i.e., 300 mW) for PLA and ABS; while only to 30 mW for PET-G (higher laser power led to irreversible damages of this sample, whereas lower laser power resulted in low signal-to-noise ratio). Automatic baseline correction was performed directly in the instrument software if not otherwise stated, any smoothing and/or corrections of the recorded RS spectra were omitted.

# 3 Results and discussion

## 3.1 Original samples

First, the mechanical properties of the original as prepared samples (before UV irradiation) will be discussed. Table 2 shows hardness (*H*) and reduced elastic modulus ($E_r$) of the original 3D printed samples without exposure to UV. A decrease in hardness and modulus of elasticity can be observed with higher indentation loads (and correspondingly greater indentation depths) for all observed samples. Such decrease is consistent with the well-documented indentation size effect (ISE) in polymers [61] further enhanced by the viscoelastic response of thermoplastic polymers which causes Oliver–Pharr analysis to overestimate modulus at shallow contact depths [62].

Table 2. Mechanical properties of original 3D printed samples obtained using nanoindentation tests under different loads. Spectroscopic measurements were performed on the grey-highlighted samples.

| | *Hardness* [GPa] | | | | *Reduced modulus* [GPa] | | | |
|---|---|---|---|---|---|---|---|---|
| | 0.5 mN | 2 mN | 10 mN | 100 mN | 0.5 mN | 2 mN | 10 mN | 100 mN |
| **PLA w** | 0.36±0.12 | 0.29±0.05 | 0.24±0.02 | 0.19±0.04 | 7.5±1.6 | 6.4±0.7 | 5.6±0.2 | 4.5±0.5 |
| **PLA w2** | 0.38±0.04 | 0.29±0.02 | 0.23±0.04 | 0.21±0.04 | 8.0±0.8 | 6.6±0.2 | 5.7±0.5 | 5.1±0.5 |
| **PLA b** | 0.34±0.10 | 0.25±0.03 | 0.24±0.08 | 0.25±0.04 | 7.4±1.5 | 6.1±0.4 | 5.8±1.0 | 5.3±0.7 |
| **ABS w** | 0.18±0.02 | 0.15±0.03 | 0.13±0.02 | 0.12±0.02 | 3.7±0.3 | 3.2±0.3 | 2.9±0.4 | 2.6±0.2 |
| **ABS b** | 0.15±0.05 | 0.12±0.04 | 0.12±0.04 | 0.11±0.02 | 3.3±0.7 | 2.9±0.5 | 2.8±0.5 | 2.4±0.2 |
| **ABST b** | 0.46±0.19 | 0.22±0.09 | 0.19±0.07 | 0.12±0.02 | 8.9±2.3 | 5.2±1.6 | 4.2±1.3 | 2.6±0.3 |
| **PETG b** | 0.20±0.07 | 0.16±0.04 | 0.13±0.04 | 0.13±0.02 | 3.9±0.6 | 3.4±0.6 | 3.0±0.4 | 2.8±0.2 |
| **PETG b2** | 0.19±0.04 | 0.16±0.02 | 0.16±0.02 | 0.13±0.02 | 3.9±0.4 | 3.5±0.3 | 3.4±0.2 | 2.9±0.3 |
| **PETG b3** | 0.26±0.04 | 0.23±0.02 | 0.22±0.04 | 0.16±0.02 | 6.0±1.1 | 6.2±1.2 | 5.8±0.6 | 4.3±0.4 |
| **HIPS w** | 0.14±0.06 | 0.14±0.02 | 0.10±0.01 | 0.10±0.02 | 3.0±1.1 | 2.9±0.3 | 2.5±0.2 | 2.2±0.3 |
| **PC w** | 0.21±0.11 | 0.20±0.09 | 0.19±0.08 | 0.21±0.08 | 4.3±1.8 | 4.4±1.6 | 4.8±1.4 | 3.7±0.9 |

The measured hardness and elastic modulus values are generally in agreement with those reported in the literature. However, since several references relate to bulk samples and not 3D printed ones, some variations can be expected due to the different preparation methods. At the same time, due to the surface effect in plastics, the deeper indentations of 10 – 100 mN loads are the most relevant for the comparison with referenced values.

Our results show that hardness of the ABS sample at given depths always corresponds to 50 % of the hardness of the PLA sample. Both samples exhibit similar trend of decrease in *H* and *E*r with depth.

Values of hardness and elastic modulus differ only slightly for black and white variants of PLA and ABS, and discrepancy is within the standard deviations. It can therefore be concluded that the mechanical properties of these two basic polymers are not affected by the additional dye. According to several sources from 3D printing community [63-65], PLA is generally considered "stronger and more durable" than ABS, which usually refers to higher tensile strength but less toughness. Several sources [66-68] confirm our finding that PLA exhibits higher surface hardness compared to ABS over the measured depth range of 250 nm to 5000 nm.

**PLA**

Hardness of PLA ranged from the maximal 0.34 – 0.38 GPa from the shallowest indentation (0.5 mN) in between three measured samples to the 0.19 – 0.25 GPa for deepest indentations (100 mN), which were in average the highest values in between all samples in this set. Cifuentes et al. [69] reports similar hardness values of extruded and moulded PLA in the range of 0.24 - 0.26 GPa, similarly to Wright-Charlesworth et al [70] with values 0.24 – 0.26 GPa. Lower hardness was reported in Batakliev et al. in the range of 0.15 – 0.17 GPa [71] which was similar to several other publications [72, 73] measured both on filaments before extrusion and samples after printing.
As for the elastic modulus, which is way less surface sensitive, our values ranged from 7.4 – 8.0 for the shallow and 4.5 – 5.3 for deepest indentations. The latter is well in agreement with values 4.0 – 5.5 GPa from Wright-Charlesworth et al [70] while beforementioned authors reported lower values 3.9 – 4.2 GPa [71-73]. These discrepancies can be attributed to a different thermoplastic manufacturing process, or to different methods of mechanical properties testing, such as tensile testing versus higher load indentation with diminished surface effect.

**ABS**

As mentioned, ABS values oh $H_{IT}$ were roughly half that of PLA, with hardness 0.15 – 0.18 GPa and 0.11 – 0.12 GPa for shallow and deep indentations, respectively. Paloma et al. [74] and Shabana et al. [10] report hardness of the ABS in the range of 0.11 - 0.13 GPa, similarly as Jyoti et. al [75] and Kapoor et. al. [76], which is consistent with our values. Higher hardness of 0.22 GPa was reported by as Bano et. al. [77]. Concerning the elastic modulus, our values ranged at 3.3 – 3.7 GPa and 2.4 – 2.6 for shallow and deepest indentations, respectively, consistent with values 2.5 – 3.5 GPa from several articles [75-77], and also to the value of 2.59 GPa stated in the "designersdata" database [78].
Rather different results were obtained for the ABST sample with twice the hardness and 12 % higher elastic modulus in comparison to the pure ABS specimen in the case of surface indentation (0.5 mN), but this difference disappears with increasing depth, while for the deepest indentation (at 100 mN) the values are already similar. It should also be mentioned that the values for shallow indentations were burdened by a significant standard deviation (up to 34 %) even after non-standard curves reduction. This is most probably caused by to low surface quality (higher roughness) of the ABST sample, complicating the measurement. It can also be the secondary effect of the methyl methacrylate additive, which may affect the oxidation resistance of the surface layer during the printing procedure.

**PETG**

The PETG samples were printed from filaments produced by three different manufacturers for more comprehensive comparison of mechanical properties. While to samples PETG b and PETG b.2 exhibited similar values for hardness end elastic modulus of $H \approx 0.20 – 0.13$ GPa and $Er \approx 3.9 – 2.8$ GPa from the shallowest to the deepest indentations, the third PETG b.3 had 30 % higher hardness and up to 54 % higher elastic modulus. The extruded PETG filament is often produced using recycled precursor, so this specific sample may have a different degree of polymerisation or even an unlisted composition. References on PETG are missing in literature, however PETG is structurally close to amorphous PET even though it is not the same. Giró-Paloma et al. [74] report PET hardness values in the vicinity of 0.25 GPa, slightly higher than our study. In contrast, the modulus of elasticity (around 3.6 GPa) is consistent with our measurements for surface indentations. Flores & Calleja [79] studied amorphous PET using depth-sensing indentation, reporting values of 0.15 – 0.25 GPa depending on depth and rate, with elastic modulus around 3 – 5 GPa. The "designerdata" database [80] states an elastic modulus of 2.95 GPa in consistency with our values for deeper indentations. Differences in mechanical property values may be caused by the different manufacturing process as different amounts of recycled component of various purity are often introduced in PET precursors [81].

**HIPS**

The HIPS sample mechanical properties were the lowest in between measured samples, with hardness $H \approx 0.14 – 0.10$ GPa from the shallow to deep indentations and modulus of elasticity $E\text{r} \approx 3.0 – 2.2$ GPa for the same depth range. This is not surprising, as the material is often use as filler or supportive material, not as the final product. References can be found mainly on basic PS (polystyrene), for instance Briscoe et al. [82] reported higher values for hardness $H \approx 0.18 – 0.28$ GPa near the surface, decreasing to 0.15 – 0.20 GPa at greater depths. These values are for the glassy PS matrix without rubber particles. The "designersdata" database [83] reports a modulus of elasticity value of 2.2 GPa for HIPS, which agrees with our values for deeper indentation.

**PC**

Polycarbonate (PC) is often described by manufacturers as one of the hardest polymeric materials available [84]. Although the measurement of the PC sample was burdened with a higher standard deviation, it ranked among the hardest of the tested samples, surpassed only by PLA and ABST. With a value of 0.2 ± 0.02 GPa, which remained constant under various loads, it exhibited the best homogeneity compared to all other samples. While modulus of elasticity of PC in the range of 3.7 – 4.8 GPa did not show a trend based on indentation depth, it was consistently around 20 % lower than for ABS. These values align well with referenced ones for bulk polycarbonate values, as it is often used as calibration reference material. Giró-Paloma et al. [74] reports similar hardness of 0.23 GPa while their elastic modulus 3.6 GPa is in our lower limit. Similar values using nanoindentation were also obtained by Briscoe et al. [82] ($H \approx 0.17–0.22$ GPa, $E\text{r} \approx 3.0–4.5$ GPa) and Iqbal [85] with hardness of 0.2 ± 0.02 GPa. The "matweb" database [86] reports an elastic modulus of 1.8 - 3.2 GPa, which is comparable to our value for the deepest indentation of 3.6 ± 0.8 GPa. It is worth noting that nanoindentation measurements of PC samples generally concern bulk materials, whereas information on 3D-printed PC using the FFF method is scarce or virtually non-existent.

## 3.2 UV lamp photometry

The UV radiation source was characterised after the exposure campaign using the same lamp type and geometry. The relative spectral power distribution was measured with a spectrometer AvaSpec-ULS2048XL-EVO by Avantes [87] and an integrating sphere over 250 – 900 nm with sub-nanometer precision. The absolute irradiance at the sample position was obtained from the photocurrent of a 1 $cm^2$ silicon photodiode by Hamamatsu S2281-01 sensors [88], converted using the effective responsivity $\eta = 0.1621$ $A.W^{-1}$, calculated by weighting the spectral responsivity of the photodiode with the measured relative spectrum of the lamp. The resulting absolute spectral irradiance is shown in Fig. 2.

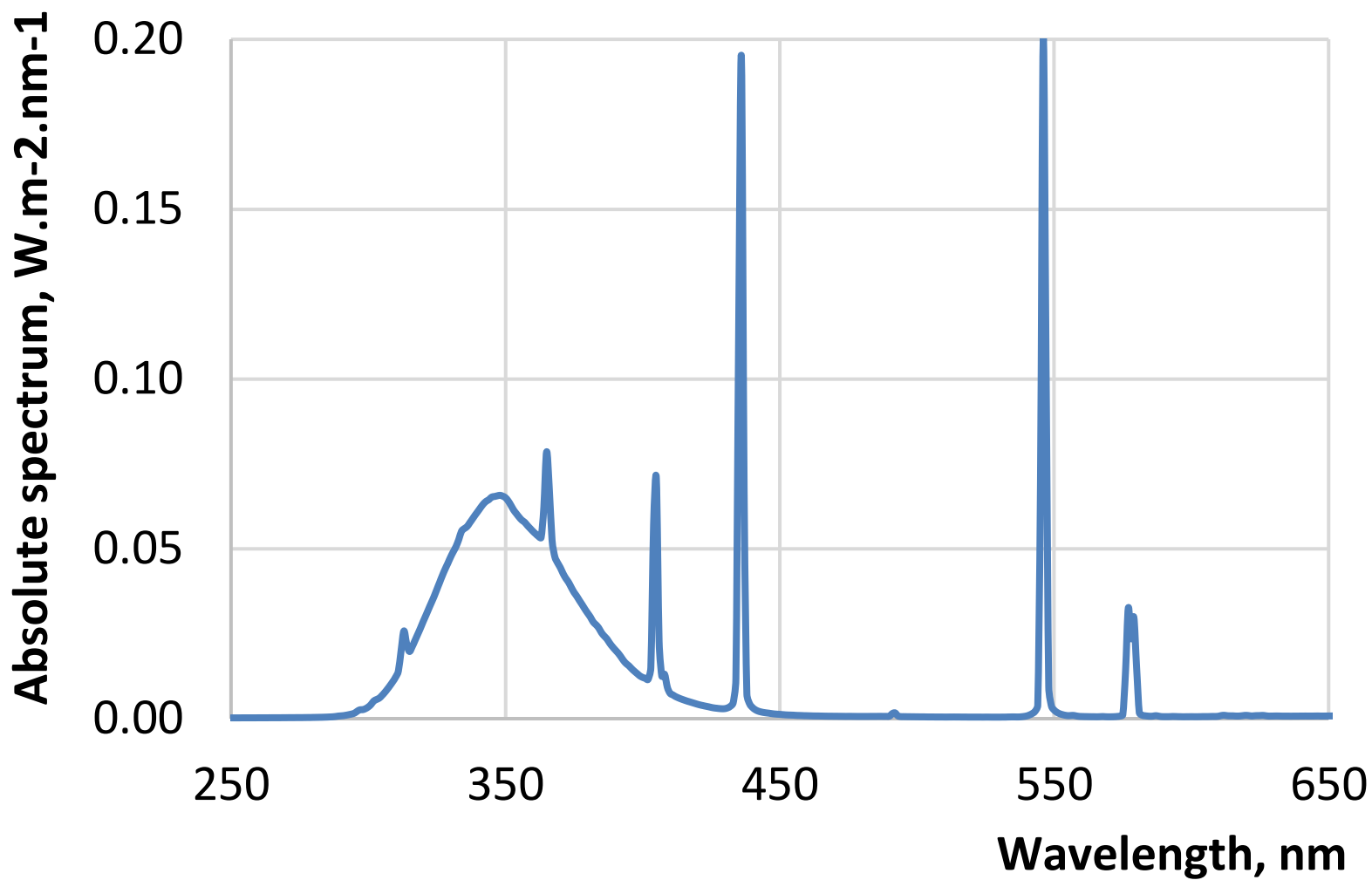

Fig. 2. Absolute spectral irradiance of the NARVA LT 18W T8/UVS-2 fluorescent lamp, measured at the specimen position (250 – 650 nm shown; integrated values in the text refer to 250 – 900 nm).

The integrated irradiance over 290 – 400 nm at the specimen position was 3.84 W.m$^{-2}$, representing 71 % of the total radiant output of the lamp between 250 and 900 nm. The spectral irradiance at 340 nm was 0.061 W.m$^{-2}$.nm$^{-1}$, approximately 15 times lower than the 0.89 W m$^{-2}$.nm$^{-1}$ specified for accelerated exposure in ASTM G154 [89]; the present exposure is therefore a long, low-irradiance one rather than an accelerated test. Emission below 290 nm was at the stray-light level of the spectroradiometer and is not considered further.

The corresponding radiant exposure was 53.5 MJ.m$^{-2}$ after 3872 h and 107.1 MJ.m$^{-2}$ after 7744 h. Taking 201 MJ.m$^{-2}$ as the annual outdoor UV dose over 290 – 400 nm in Central Europe [90], these correspond to approximately 0.27 and 0.53 years of outdoor UV exposure, respectively. The equivalence is, however, strongly band-dependent (see Table 3) – relative to sunlight at equal total UV dose, the exposure over-represents the region below 360 nm and under-represents the longer UV-A, so that the 7744 h exposure delivered about three quarters of a year of outdoor dose below 360 nm but only about one third of a year above it. Nevertheless, this is relevant to the interpretation of the material-dependent results, since the aromatic-backbone polymers in this study absorb predominantly at shorter wavelengths [16, 23, 26], where the exposure was comparatively accelerated.

Irradiance uniformity over the sample holder was verified by measurement at six positions 0, 2.5 and 5 cm in longitudinal axis and 0 and 2.5 cm in perpendicular axis from the central position below the lamp. All positions received at least 96 % of the maximum value, i.e. a spread below 4 %, which is negligible compared with the scatter of the measured mechanical properties.

Table 3. Absolute spectral irradiance of the NARVA LT 18W T8/UVS-2 fluorescent lamp, measured at the specimen position (250–650 nm shown; integrated values in the text refer to 250 – 900 nm).

| Spectral band | Irradiance *E* [W m$^{-2}$] | Share of 290-400 nm [%] | *H* at 3872 h [MJ.m$^{-2}$] | *H* at 7744 h [MJ.m$^{-2}$] | Annual outdoor solar dose [MJ.m$^{-2}$] | Equivalent outdoor years at 7744 h * |
|---|---|---|---|---|---|---|
| **290 – 320 nm** | 0.32 | 8.4 | 4.5 | 9.0 | 11.9 | 0.76 |
| **320 – 360 nm** | 2.19 | 56.9 | 30.5 | 61.0 | 81.1 | 0.75 |
| **360 – 400 nm** | 1.33 | 34.7 | 18.6 | 37.1 | 108.0 | 0.34 |
| **290 – 400 nm** | 3.84 | 100 | 53.5 | 107.1 | 201.0 | 0.53 |

*Equivalent outdoor years refer **exclusively to the accumulated UV component** of outdoor exposure. They do not account for temperature cycling, humidity, condensation or mechanical stress, and are not predictions of outdoor service life.

As the characterisation was performed after the exposure campaign, the reported values correspond to a lamp past its initial burn-in rather than to a new one. For fluorescent UV lamps the radiant output decreases with accumulated operating time, while the spectral power distribution remains essentially unchanged [91, 92]; the manufacturer specifies a lamp maintenance factor of 89 % at the end of the 20 000 h rated life [93]. The spectral shares given in Table 3 and the comparison in Fig. 2 are therefore insensitive to lamp age, whereas the absolute irradiance carries a corresponding uncertainty of the order of 10 %. The chamber had no irradiance feedback control, and no lamp replacement schedule was applied during the exposure.

## 3.3 UV exposed samples

The main aim of the paper was to study the photooxidative degradation of polymer samples after exposure to continuous UV irradiation. The other parameters - temperature and humidity - were kept constant and of low values at the same time in order to separate the effect of photooxidation. The changes in mechanical properties, especially their depth-dependence, and the correlation of these changes to structural changes were studied.

### 3.3.1 Surface structural changes

One of the most apparent effects of long-term UV exposure on a polymer is so-called discoloration or decolourization caused by fading pigment, which can be observed even by the naked eye. The decolourization has been previously noted by several authors [33, 36, 94]. Within the studied set of samples, this effect can be observed after UV exposure to 3872 hours specifically on white PC, HIPS and ABS samples that turned to yellow-grey colour (see Fig. 3). Interestingly, further UV irradiation up to 7744 hours, did not lead to further change in colour. This is in a reasonable agreement with the literature. For instance, in Davis et al. [33], the decolourization is explained as the formation of photoproduced chromophores during ABS degradation (photo-oxidation). The as-formed chromophores absorb energy in the UV-visible spectrum and thus cause the yellowish colour of the polymer after UV exposure.

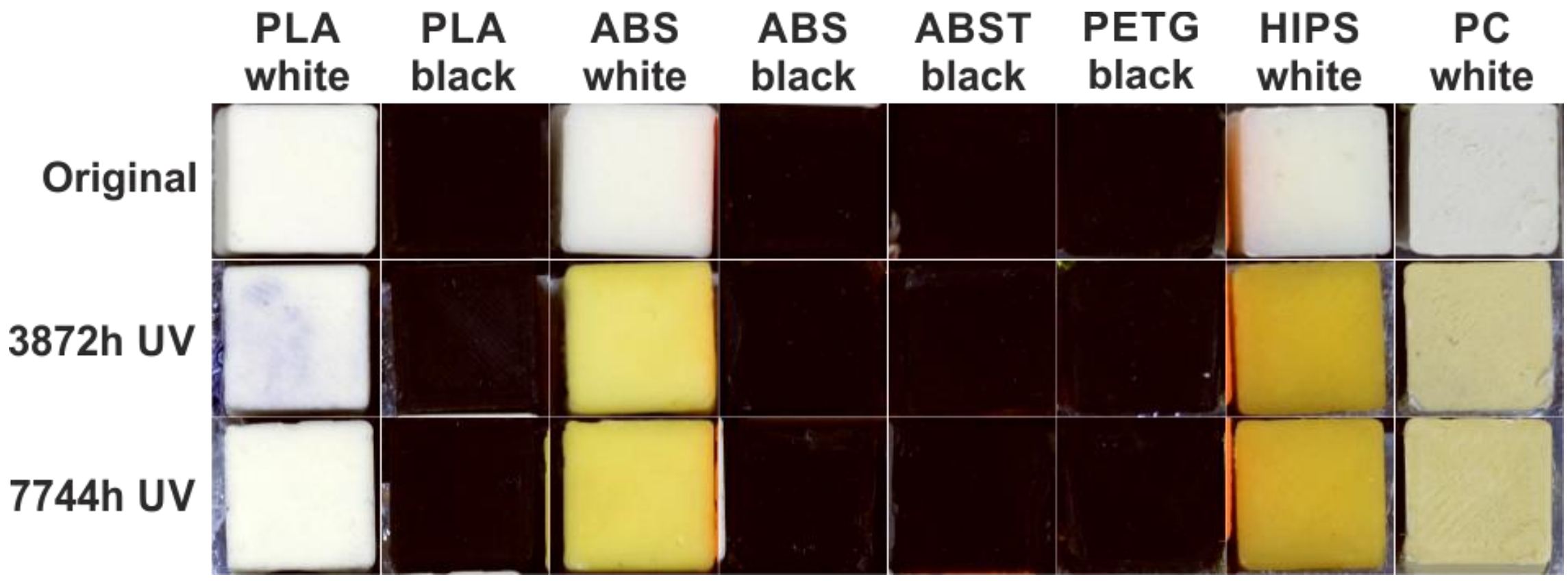

Fig. 3. Samples of 3D printed polymers in forms of 1x1x1 cm cubes used in this study. Note the decolourization after UV exposure in second and third lines.

Further detailed examination of the surfaces was performed by laser scanning confocal microscopy, which allows the specimen to be observed without the conductive coating required for SEM and therefore preserves the original surface. Fig. 4 compares five of the studied materials: the left column shows unexposed samples from the control set, the middle column specimens after 3872 h and the right column after 7744 h of UV exposure. The dominant texture in all images is the as-printed surface of the first deposited layer, formed against the heated build plate.

The materials differ markedly in their response. PLA w (see Fig. 4a) shows no cracking at any exposure time; only the rounded features of the as-printed texture and isolated particles are present. ABS w (see Fig. 4b) develops individual long cracks traversing the entire field of view, already after 3872 h, but these remain isolated and do not form a network. PETG b, HIPS w and PC w (Fig. 4c-e) instead develop interconnected crack networks. The progression is clearest for HIPS w (see Fig. 4d) with a polygonal network with a cell size of the order of 100 µm after 3872 h develops into a markedly finer and strongly branched network with a cell size of the order of 20 µm after 7744 h, and cracks are frequently seen to originate at the pre-existing surface voids visible in the unexposed specimen. In PETG b a fine network is already present after 3872 h, whereas after 7744 h fewer but longer and more widely opened cracks with T-shaped junctions dominate. In PC w isolated fine linear features can be discerned after 3872 h, but a clearly developed branched network is apparent only after 7744 h. Since the three columns correspond to different specimens and different fields of view, the comparison is qualitative. It nevertheless provides direct morphological evidence of brittle surface damage that is independent of the indentation response.

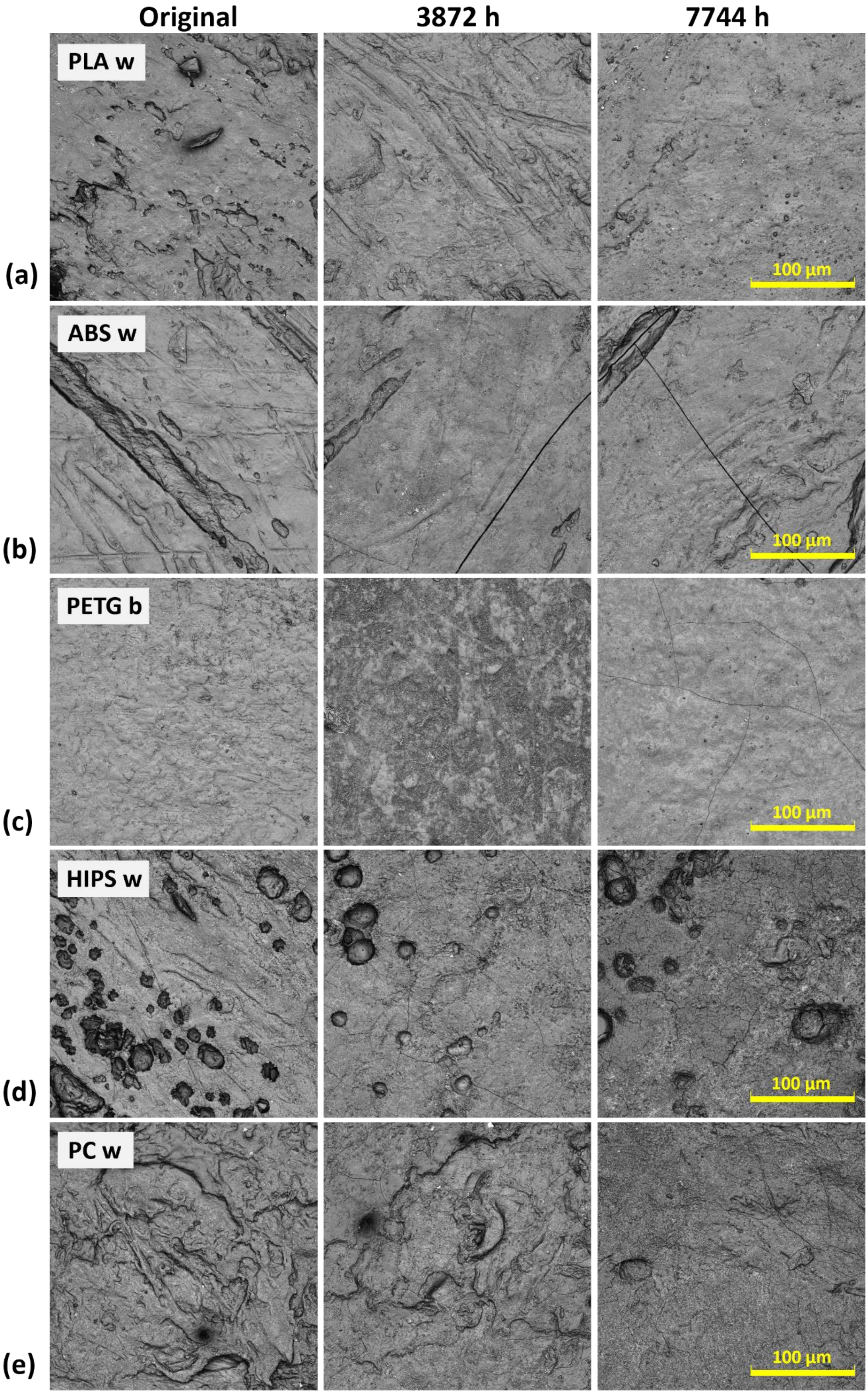
Original
3872 h
7744 h
(a)
PLA w
100 µm
(b)
ABS w
100 µm
(c)
PETG b
100 µm
(d)
HIPS w
100 µm
(e)
PC w
100 µm

Fig. 4. Laser scanning confocal micrographs of the samples surfaces: (a) PLA w, (b) ABS w, (c) PETG b, (d) HIPS w, (e) PC w. Left column – unexposed control specimens; middle column – after 3872 h; right column – after 7744 h of UV exposure. Field of view 256 x 256 µm.

Selected samples exposed to UV for 7744 h were additionally examined by scanning electron microscopy, which resolves the crack opening and the junction morphology at the micrometre scale that the confocal measurement cannot. The specimens were sputter-coated with a thin conductive layer of Cu, and the examination was therefore performed only after the confocal measurements had been completed. Fig. 5 compares PETG b (a, b) with PC w (c, d). In PETG the cracks are few but long and widely opened, and terminate against one another in T-shaped junctions (Fig. 5b), a morphology typically associated with sequential rather than simultaneous crack formation. In PC the network is hierarchical, with a small number of wide primary cracks and a denser population of finer secondary cracks running approximately perpendicular to them and delimiting cells of the order of 10 µm (Fig. 5d). Both observations are consistent with the confocal results described above.

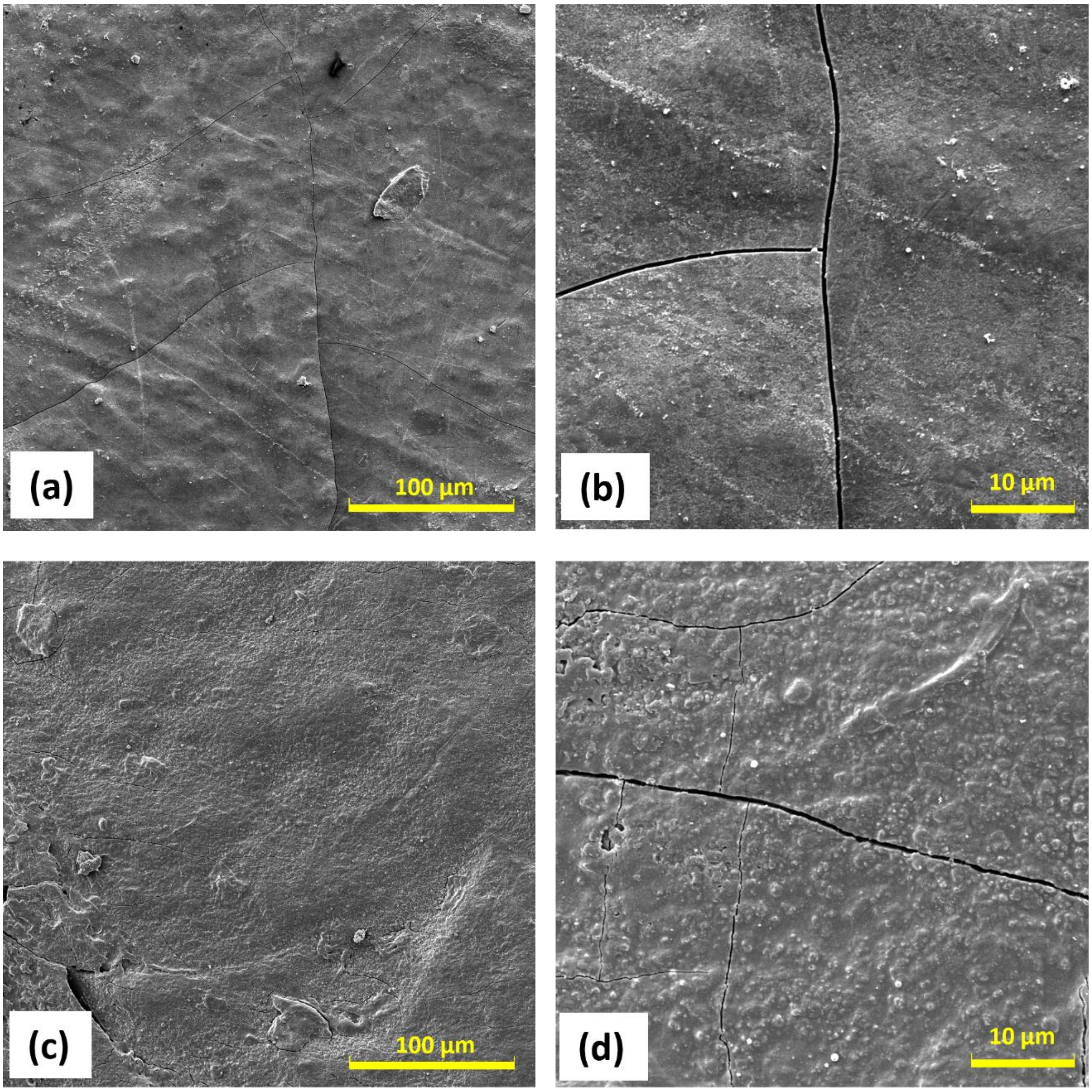

Fig. 5. Scanning electron micrographs of samples exposed for 7744 h: (a, b) PETG black and (c, d) PC white samples in two magnifications – 1280x on the left, 8000x on the right.

### 3.3.2 Mechanical properties

The changes in mechanical properties resulting from UV exposure are summarised in Fig. 6 and Fig. 7, which show eight of the eleven samples of Table 1 at three of the four loads used. The omitted samples – PLA w2, PETG b2 and PETG b3 – behaved after exposure as their counterparts PLA w and PETG b; PETG b3, which differed in the original state, became similar to the other PETG samples after long-term exposure. Complete results for all samples are given in Fig. SI2 and Fig. SI3.

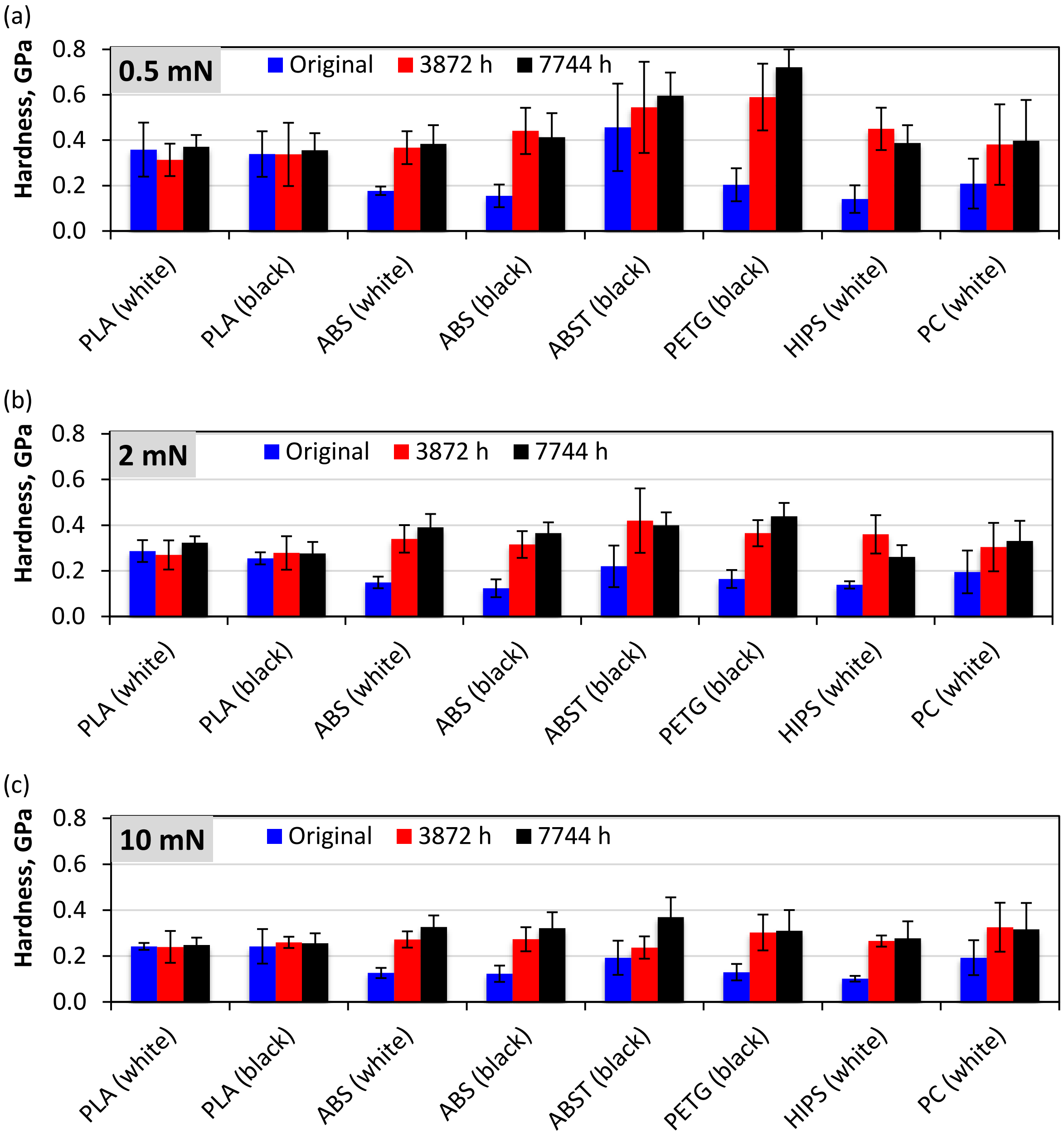

Fig. 6. Nanoindentation hardness $H_{IT}$ for selection of samples from Table 1 comparing values for original state, 3872 hours and 7744 hours of summarized UV exposure obtained for load of (a) 0.5 mN, (b) 2 mN, (c) 10 mN.

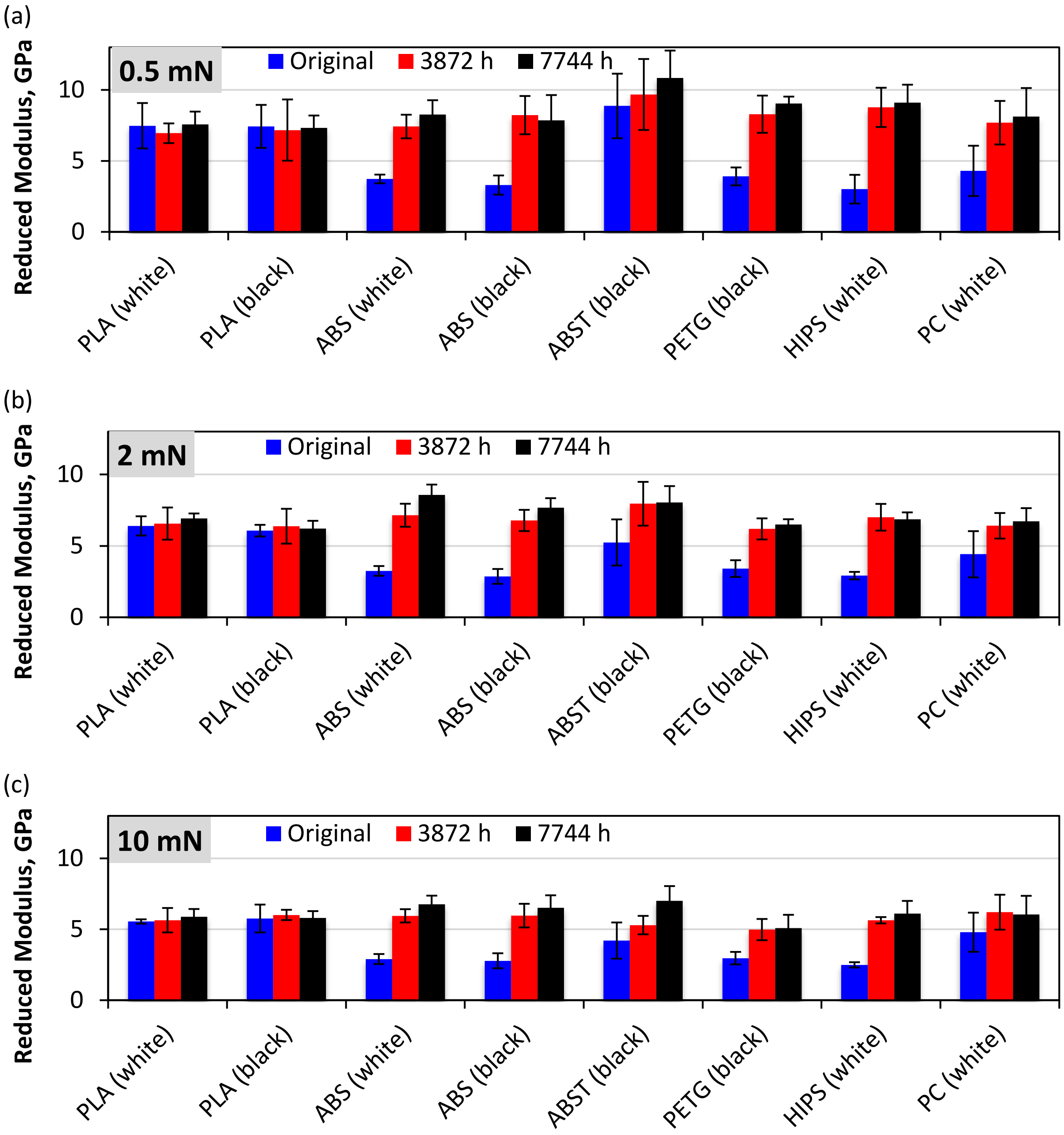

Fig. 7. Nanoindentation reduced elastic modulus $E_r$ for selection of samples from Table 1 comparing values for original state, 3872 hours and 7744 hours of summarized UV exposure obtained for load of (a) 0.5 mN, (b) 2 mN, (c) 10 mN.

Both hardness and reduced modulus increase with UV exposure for all materials except PLA. An increase in hardness can improve wear resistance but is also commonly associated with increased brittleness [95-97], and specific fragility parameters that rise with hardness have been proposed [98, 99].

The most surface-sensitive data are those at 0.5 mN, reaching depths of 150–200 nm, but these are also the most affected by surface irregularities and show the largest standard deviations; the 2 mN load, reaching 400 – 500 nm, is the best compromise for monitoring surface changes. At 0.5 mN, only two PLA samples and ABS-T exceeded 0.3 GPa in hardness and 6 GPa in reduced modulus in the original state, whereas after 3872 h all eight samples exceeded both values. The same trend is seen at 2 mN with lower absolute values, reflecting the larger contribution of the unaffected bulk at greater depth. The magnitude of the change differs considerably between materials.

Furthermore, the samples will be discussed individually:

**PLA after UV treatment**

PLA, represented by three samples, showed no measurable change in either quantity even after 7744 h; the averages remained within the standard deviation at all loads. This is notable, since PLA is generally regarded as the least durable of the materials studied, its photo-oxidation proceeding by random chain scission to carboxylic acid and diketone end groups [48]. Additional PLA samples containing 1 – 3 vol.% of a UV stabiliser gave results identical to the unstabilised material (see Fig. SI4).

**ABS after UV treatment**

In contrast, ABS, the second of the most used materials for 3D printing, changed markedly. Hardness increased from 0.17 to almost 0.40 GPa at 0.5 mN and from 0.11 to 0.22 GPa at the deepest indentations, for both the white and the black variant, and the reduced modulus doubled over the corresponding depths. Whereas ABS originally reached about half the hardness and modulus of PLA, after 3872 h it exceeded PLA in both, and more so after 7744 h. This agrees with Davis et al. [33] reporting a loss of impact strength attributed to densification and embrittlement of a surface layer. The mechanism is photolysis of methylene bonds in the polybutadiene phase, producing free radicals that generate carbonyl and hydroxyl groups and crosslinking [33, 49]; the white ABS also discoloured to a yellow-brown shade (Fig. 3). ABS-T behaved similarly, with deeper indentations giving values close to those of ABS (see Fig. 3).

**PETG after UV treatment**

PETG showed a threefold increase in hardness at 0.5 mN and a twofold increase at 2 mN, consistent with Blais et al. [26], who report increasing brittleness correlated with increasing hardness after 600 h of UV exposure.

**HIPS after UV treatment**

HIPS behaved similarly to PETG, with an almost threefold increase in hardness at 0.5 and 2 mN and more than a twofold increase at 10 and 100 mN, accompanied by pronounced discoloration (Fig. 3). Prasad et al. [50] observed changes in HIPS thin films after 50 h of UV exposure, corresponding to complete oxidation of the polybutadiene phase, with little further change thereafter, which agrees with our results. One possible explanation is the formation of a surface layer limiting oxygen access, so that oxidation continues at a reduced rate. This is not demonstrated by the present data, and a limited penetration depth of the UV radiation, depletion of photo-reactive species or stabilisers, changes in the optical absorption of the degraded surface, or the limited probing depth of the techniques used are equally consistent with them. A similar levelling-off is seen to some extent in all the materials examined.

**PC after UV treatment**

PC showed an increase of approximately 80 % in hardness and reduced modulus at 0.5 mN after 3872 h, less pronounced at greater depths, with little further change up to 7744 h. PC is generally regarded as susceptible to UV exposure [51, 52]. Sheerman et al. [53] reported the formation of a brittle surface layer that may spall and thereby arrest crack propagation into the bulk, while Ramani et al. [23] found degradation in both ABS and PC but considerably more in ABS, in agreement with our results. PC degradation is thus a superficial phenomenon extending a few micrometres below the surface, which does not usually alter bulk properties [54] but is detectable by the locally sensitive nanoindentation technique.

### 3.3.3 Load-partial unload tests - Hardness and elastic modulus depth profiles

Hardness and reduced modulus depth profiling was used to explore the extent of the surface modification caused by UV exposure. The load-partial unload (LPU) regime of nanoindentation provides this profile from a single indentation site and is therefore suited to visualising the depth over which the structure has been altered.

The depth-resolved indentation response originates from a deformation volume extending below and laterally around the contact, so a value obtained at a given depth reflects that surrounding volume rather than a plane at exactly that depth. The LPU results are therefore used to demonstrate the presence and extent of a mechanical-property gradient, not to determine the exact thickness of a degraded layer.

The standard nanoindentation results above were obtained from matrices of indentations at different locations; because FFF produces some shape inhomogeneity of the surface, part of those results carry a higher standard deviation. LPU indentation at a single spot gives a more sensitive probe of the properties over a range of depths. The load was segmented into 30 cycles between 5 and 200 mN, giving indentation depths from 800 to 8500 nm. The initial load of 5 mN is higher than the 0.5 and 2 mN used in the standard tests, so the LPU data are less surface-sensitive. The results are plotted as hardness (Fig. 8) and modulus of elasticity (Fig. 9) against depth below the surface.

PLA (Fig. 8a, Fig. 9a) showed stable hardness and modulus across the whole range from about 900 to 6700 nm, for the original sample and for both exposure steps. This confirms the standard indentation result that PLA resists photo-oxidation even at the surface. The slightly higher value at the first point is attributed to surface inhomogeneity, as it appears equally in the original and the exposed samples.

ABS (Fig. 8b, Fig. 9b) changed significantly, with a marked depth dependence. Both hardness $H$ and reduced modulus $E$r are highest at the surface and decrease with increasing depth. This trend is weakly present in the original sample, where it may reflect the indentation size effect, but is much more pronounced in the UV-exposed ones, for which the difference between the shallowest (≈ 0.9 µm) and the deepest (≈ 7 µm) indentations is around 57 – 75 % for both hardness and modulus in both UV exposures. Even at the deepest indentations, however, the values of the original unexposed sample are not reached. HIPS (Fig. 8d, Fig. 9d) follows the same trend.

PETG (Fig. 8c, Fig. 9c) behaves differently: the elevated values are confined to a surface layer and fall back to those of the unexposed sample at about 5 µm. The change in PETG is therefore more surface-confined than in ABS or HIPS, in agreement with the standard indentation data, which probed as shallow as 300 nm at 0.5 mN and showed PETG to be the most strongly affected material.

PC (Fig. 8e, Fig. 9e) appears essentially stable after exposure, most clearly in the reduced modulus. The hardness curves vary somewhat, but much less than in ABS, PETG and HIPS – comparably to PLA, where no effect was found – and they converge beyond about 4.5 µm. Since the PC profile begins at a relatively large depth, this variation more likely reflects scatter than a real gradient, consistent with the standard indentation result that the effect on PC was measurable but confined to a few micrometres.

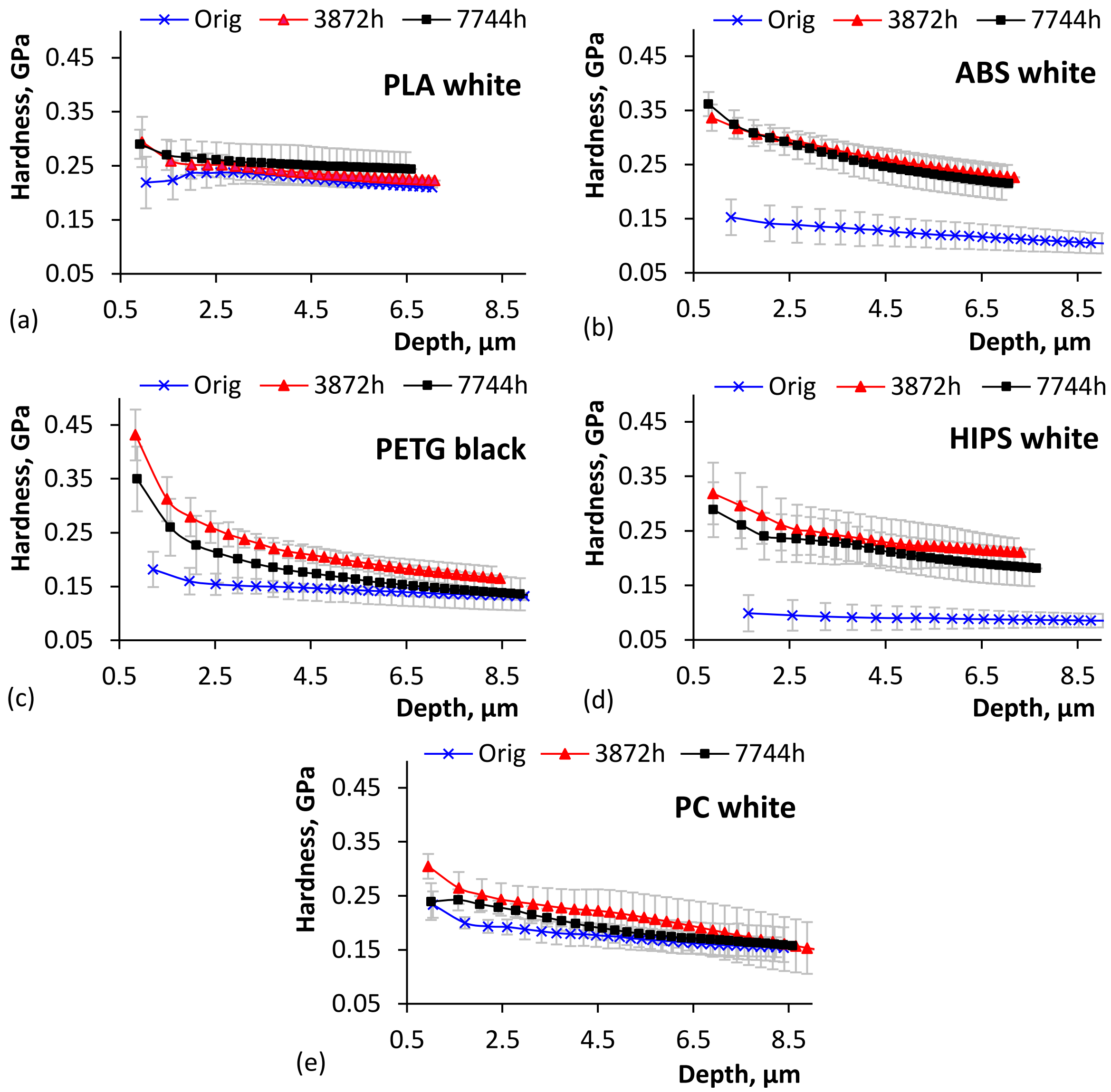

Fig. 8. Depth profiles of nanoindentation hardness from load-partial unload (LPU) tests for (a) PLA w, (b) ABS w, (c) PETG b, (d) HIPS w and (e) PC w samples comparing original states and UV exposure after 3872 h and 7744 h.

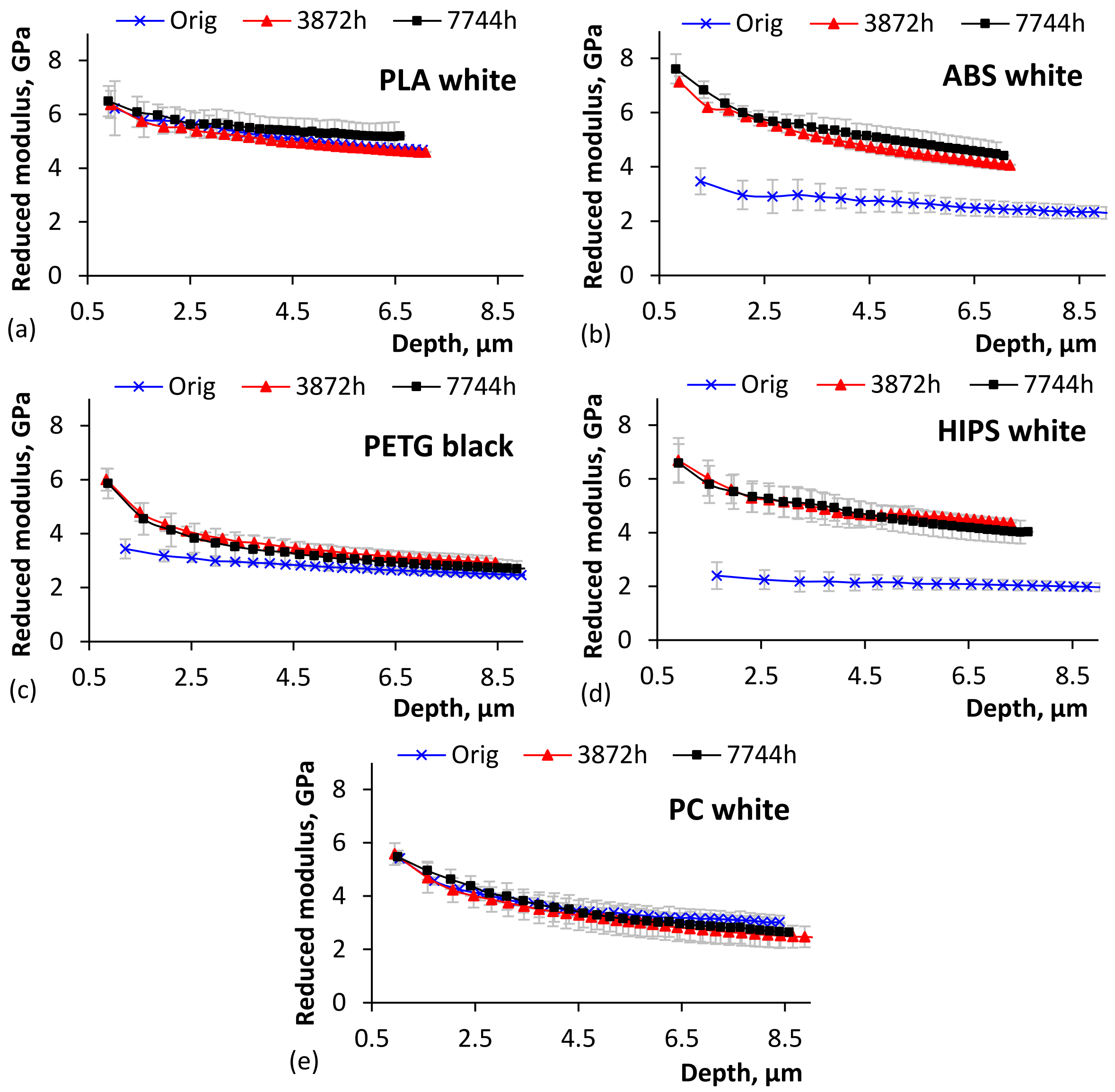

Fig. 9. Depth profiles of nanoindentation reduced elastic modulus from load-partial unload (LPU) tests for (a) PLA w, (b) ABS w, (c) PETG b, (d) HIPS w and (e) PC w samples comparing original states and UV exposure after 3872 h and 7744 h.

### 3.3.3 Creep evaluation

As described in the Experimental section, three parameters describing creep were evaluated (Table 4, Fig. 10). The differences are largest between materials, with a systematic effect of UV exposure in those already identified as UV-sensitive.

Table 4. Creep results of all tested polymeric samples under 20 mN load, evaluated using three creep parameters.

| | Curve slope [nm/s] | | | Relative depth difference [%] | | | $P_{CR}$ [nm²/s] | | |
|---|---|---|---|---|---|---|---|---|---|
| | Original samples | 3872 h UV | 7744 h UV | Original samples | 3872 h UV | 7744 h UV | Original samples | 3872 h UV | 7744 h UV |
| **PLA white** | 0.14±0.02 | 0.16±0.02 | 0.14±0.01 | 12.0±0.8 | 13.8±1.1 | 13.5±0.7 | 33±6 | 40±7 | 35±5 |
| **PLA black** | 0.15±0.01 | 0.14±0.02 | 0.13±0.03 | 13.5±1.0 | 13.1±0.7 | 12.0±1.7 | 38±6 | 33±7 | 30±10 |
| **PLA white,b** | 0.10±0.03 | 0.10±0.03 | | 12.4±1.4 | 12.1±0.3 | | 25±8 | 22±9 | |
| **ABS white** | 0.26±0.02 | 0.18±0.02 | 0.16±0.02 | 16.0±1.0 | 14.8±1.1 | 13.9±1.1 | 106±15 | 44±8 | 36±7 |
| **ABS black** | 0.23±0.03 | 0.16±0.01 | 0.16±0.03 | 14.4±0.9 | 15.1±0.7 | 16.0±1.4 | 85±18 | 39±2 | 37±8 |
| **ABST black** | 0.23±0.03 | 0.13±0.01 | 0.15±0.02 | 14.5±2.3 | 12.2±0.9 | 14.1±1.3 | 84±22 | 28±4 | 36±8 |
| **PETG black** | 0.15±0.02 | 0.13±0.01 | 0.09±0.01 | 10.4±0.7 | 10.6±1.0 | 9.3±0.9 | 36±7 | 26±3 | 15±3 |
| **PETG black,b** | 0.14±0.03 | 0.12±0.01 | | 10.4±1.7 | 9.9±0.6 | | 50±14 | 22±3 | |
| **PETG black,c** | 0.12±0.01 | 0.10±0.01 | | 11.3±0.7 | 10.2±0.9 | | 31±7 | 18±2 | |
| **HIPS white** | 0.25±0.04 | 0.12±0.02 | 0.13±0.03 | 12.6±1.2 | 11.0±0.7 | 11.2±1.5 | 94±19 | 24±8 | 29±11 |
| **PC white** | 0.12±0.01 | 0.10±0.02 | 0.07±0.02 | 8.9±0.6 | 11.5±1.8 | 9.1±0.8 | 22±5 | 21±5 | 11±3 |

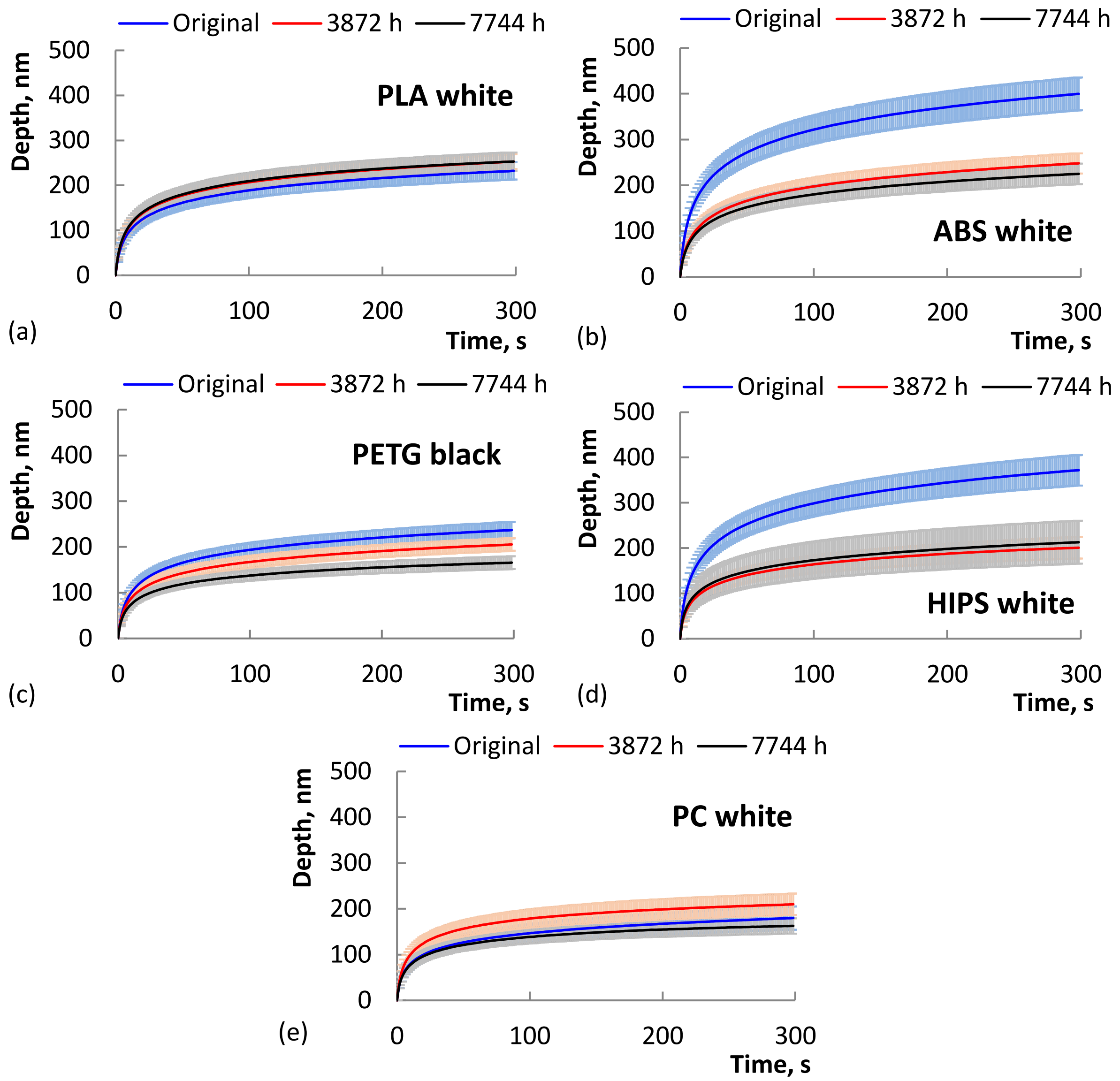

Fig. 10. Comparison of the creep curves for the Original, 3872 h and 7744 h UV exposed sets of (a) PLA w, (b) ABS w, (c) PETG b, (d) HIPS w and (e) PC w polymeric samples.

Creep of the three PLA samples was unchanged by exposure, as was the case for standard nanoindentation and LPU. Among the original samples, PC showed the lowest creep – relative depth difference 8.9 ± 0.6 %, $P_{CR}$ = 22 ± 5 nm$^2$/s, and the second lowest in curve slope 0.12 ± 0.01 nm/s - followed by PETG, whereas ABS and HIPS crept considerably more, with curve slopes of 0.23 – 0.26 nm/s and $P_{CR}$ in the range 84 – 106 nm$^2$/s.

These differences are much less pronounced in the relative depth difference, which suggests a limited resolution of this ISO-standardised parameter [58] for samples whose creep differs only slightly in compression-based indentation.

UV exposure markedly reduced the creep of ABS and HIPS: the curve slope fell from 0.23 – 0.26 in the original state to 0.12 – 0.18 after 3872 h, with a corresponding decrease in $P_{CR}$ to 24 – 44 nm$^2$/s in the same samples. This is consistent with the increase in hardness, and the associated embrittlement, found for the same materials. PETG also crept less after exposure, though to a smaller extent and mainly in the $P_{CR}$ parameter (see Table 4).

Further UV irradiation up to 7744 hours resulted in a minor change in creep $P_{CR}$ parameter for PETG b and PC w samples, while the remaining samples were stable within scatter (Table 4, Fig. 10).

The creep parameters are used here as comparative indicators under the defined experimental conditions and allow the materials and exposure states to be ranked against one another. A 300 s indentation creep experiment does not establish long-term macroscopic creep behaviour, and the results should not be extrapolated to long-term performance.

### 3.3.4 Structural changes investigated by vibrational spectroscopies

Vibrational spectroscopies (Infrared-IR absorption and Raman spectroscopy-RS) were exploited for the investigation of structural changes induced by the long-term UV irradiation in selected samples, namely, PLA, ABS, PET-G. These three samples were selected based on their different changes in mechanical properties after UV irradiation discussed in previous sections. Both spectroscopic techniques provided valuable insights and vibrational spectra. Three samples of each polymer were mutually compared: original, after 3872 hours and after 7744 hours UV exposures. The appropriate vibrational spectra for these polymers are shown in Figures 11-13, band assignments are listed in Tables 5-7, further details are presented in **Supporting Information** as Fig. SI8, SI9 and Tables SI1 and SI2.

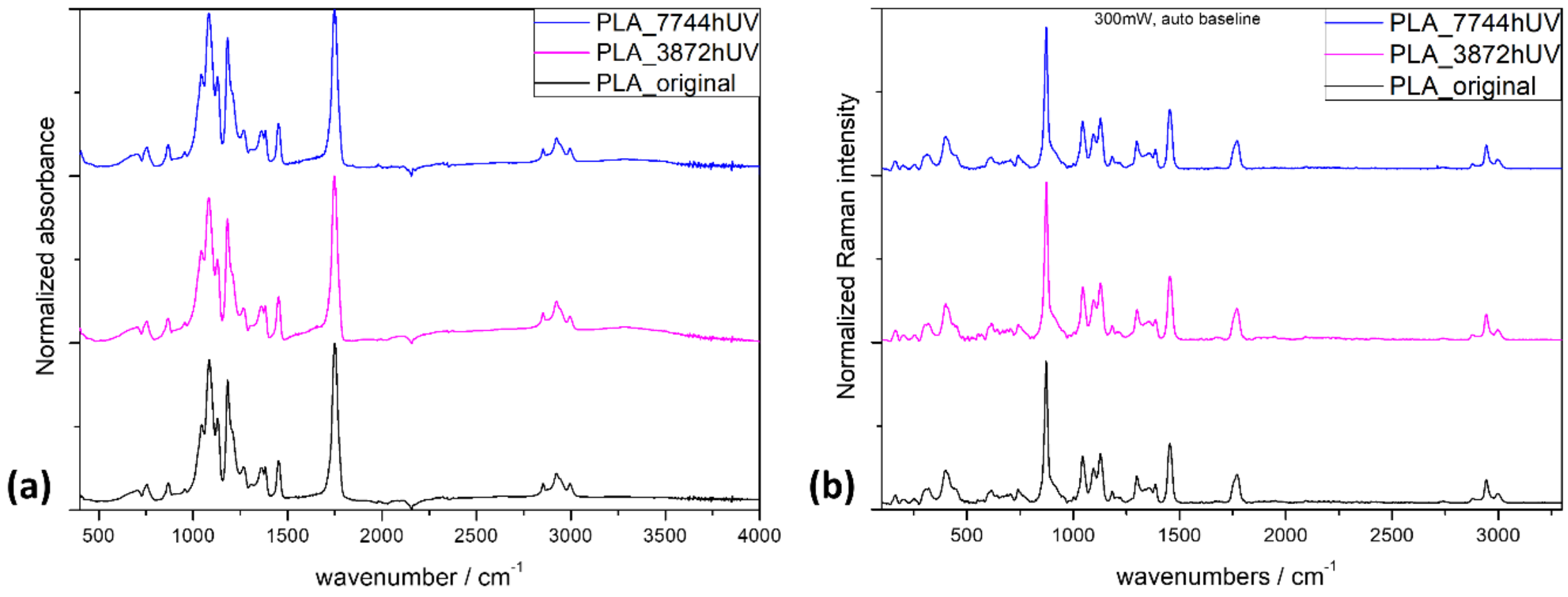

Fig. 11. Vibrational spectra of PLA: (a) IR and (b) Raman spectra. Spectra were normalized to the most intense peak at 1749 cm$^{-1}$ for IR and at 871 cm$^{-1}$ for RS. Raman spectra were obtained using the 785 nm excitation laser wavelength. Normalized absorbance in IR and normalized Raman intensity in RS spectra are presented as functions of increasing wavenumber for the sake of their direct comparison.

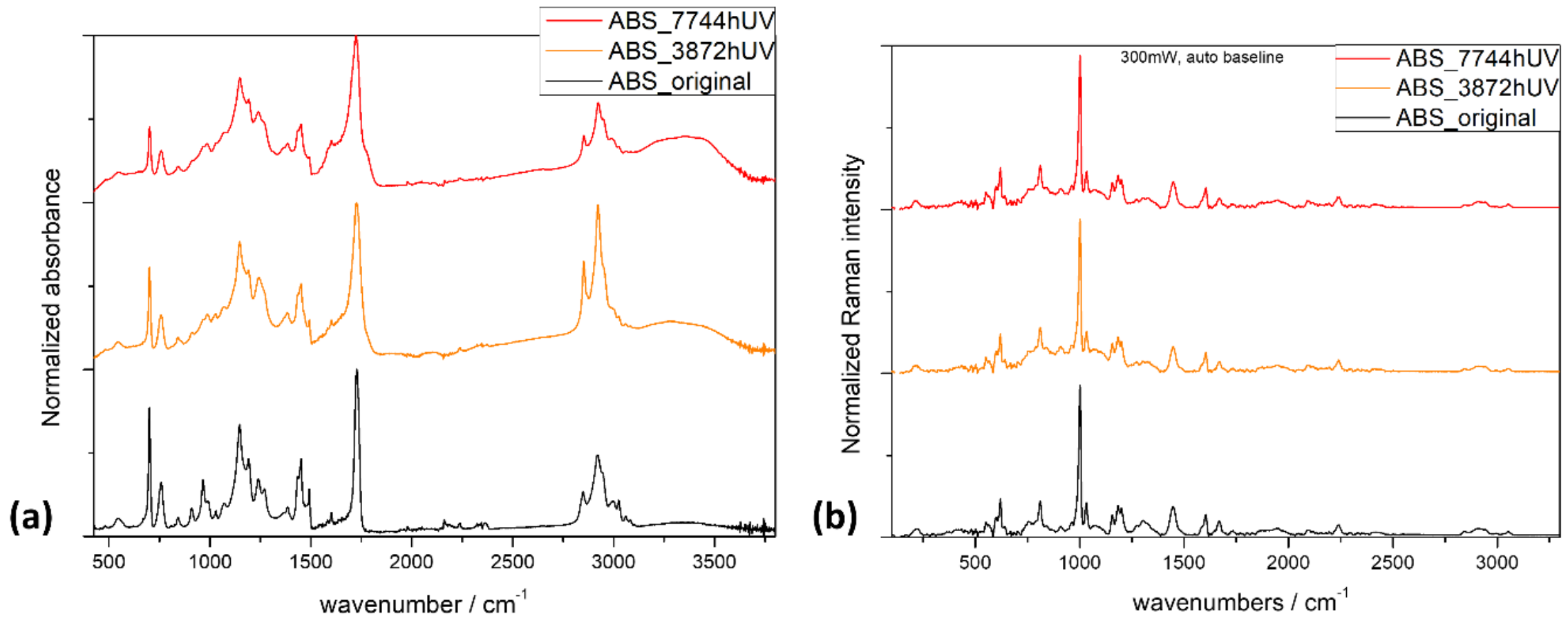

Fig. 12. Vibrational spectra of ABS: (a) IR and (b) Raman. Spectra were normalized to the most intense peak located at 1727 cm$^{-1}$ for IR and at 1002 cm$^{-1}$ for RS. Raman spectra were obtained using the 785 nm excitation laser wavelength. Normalized absorbance in IR and normalized Raman intensity in RS spectra are presented as functions of increasing wavenumber for the sake of their direct comparison.

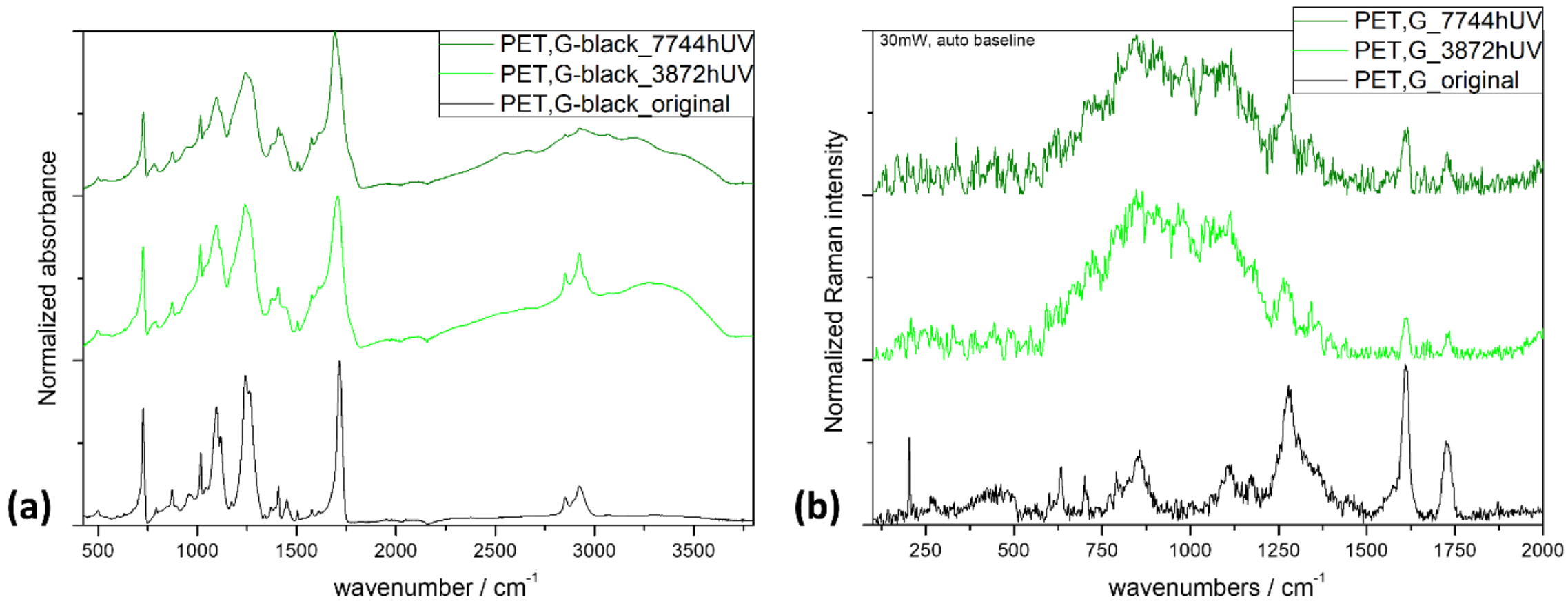

Fig. 13. Vibrational spectra of PET-G samples: a) IR and b) Raman. IR spectra were normalized to the most intense peak at 1715 cm$^{-1}$; while RS spectra to the most intense peak after automatic baseline correction. Raman spectra were obtained using the 785 nm excitation laser wavelength. Normalized absorbance in IR and normalized Raman intensity in RS spectra are presented as functions of increasing wavenumber for the sake of their direct comparison.

Table 5. IR and RS intensive bands distinguished in PLA spectra (before and after long-term UV exposure, being both identical) and their vibrational assignment (according to [43, 100]).

| IR / cm$^{-1}$ | band assignment | RS / cm$^{-1}$ | band assignment |
|---|---|---|---|
| 2993<br>2921+sh2940 | C-H stretching | 2996<br>2944+sh2940<br>2880 | C-H stretching of $CH_3$ and -$CH_2$- groups |
| 2852 | methylene (-$CH_2$-) stretching | | |
| | | 1770+sh1758 | C=O stretching |
| 1749 | C=O stretching | | |
| 1452 | $CH_3$ bending | 1454 | $CH_3$ asymmetric deformation |
| 1381 | CH bending | 1385 | $CH_3$ bending |
| 1360 | CH bending | 1360<br>1352 | CH out-of-plane bending |
| | | 1298 | CH in-plane bending |
| 1267 | $CH_3$ bending, C-O stretching | | |
| 1182+sh1205 | C-O-C stretching | 1182 | |
| 1128 | $CH_3$ rocking | 1128 | $CH_3$ bending |
| | | 1094 | C-O-C stretching and/or $CH_3$ out-of-plane deformation |
| 1084 | symetric C-O-C stretching | | |
| 1043 | C-$CH_3$ stretching | 1044 | C-$CH_3$ stretching |
| 956 | skeletal band | | |
| 867<br>755 | somehow related to amorphous and crystalline phase within PLA and C-C stretching | 871<br>742+sh758 | C-COO stretching |
| | | 612<br>400+sh444<br>306 | |

Table 6. IR and RS intensive bands distinguished in ABS spectra and their vibrational assignment (based on [43, 100])

| IR / cm$^{-1}$ | assignment | RS / cm$^{-1}$ | assignment |
|---|---|---|---|
| 3100-3600 | O-H stretching | | |
| 3025 | stretching of =C-H | 3048-3054 | stretching of =C-H |
| 2995<br>2991+sh2948 | stretching of aliphatic -C-H | 2986-2996<br>2944 and 2906 | C-H stretching of $CH_3$ and -$CH_2$- groups |
| 2849 | methylene stretching | 2840 | |
| 2237 | C≡N stretching | 2238 | C≡N stretching |
| | | 1770+sh1758 | C=O stretching |
| 1727 | C=O stretching[b] | | |
| | | 1668 | C=C stretching of butadiene |
| 1601 | styrene aromatic ring (aromatic C=C) | 1604 | C-C stretching of benzene ring from styrene |
| 1492 | aromatic ring of styrene (aromatic C=C) | | |
| 1452 | $CH_2$ scissoring mode | 1448 | $CH_2$ vibration |
| 1434 | | | |
| 1385 | CH bending | | |
| 1360 | CH bending | | |
| | | 1304 + sh1334 | C-C stretching and/or C-C-H deformation |
| 1267 | | 1274 | |
| 1241 | | | |
| | | 1200 | C-C stretching |
| 1192 | | 1184 | in-plane deformation of CH in styrene |
| | | 1156 | out-of-plane deformation of CH in styrene |
| 1146 | | | |
| 1128 | $CH_3$ rocking | | |
| 1069 | | | |
| 1029 | | 1032 | CH in-plane vibration of benzene ring in styrene |
| | | 1002 | benzene ring-breathing in styrene |
| 990 | | | |
| 965 | C=C-H deformation of 1,4-butadiene[a] | | |
| 909 | C=C-H deformation of 1,2-butadiene[a] | | |
| 840 | | | |
| | | 810 | |
| 757 | CH deformation | | |
| 699 | out-of-plane CH vibration of styren aromatic ring | | |
| | | 620 | |

Note: [a]bands assignment based on [36] as well; [b]an additive/impurity in original ABS

Table 7. IR and RS intensive bands distinguished in PET spectra and their vibrational assignment (RS based on [39]; IR based on [34, 38, 100-102]).

| IR / $cm^{-1}$ | assignment | RS / $cm^{-1}$ | assignment |
|---|---|---|---|
| 3180-3530 | O-H stretching | | |
| 3430[c] | overtone of C=O stretching of carboxylic acid | | |
| 3059, 3068, 3077 | stretching of =C-H | | |
| sh2954[b]<br>2922 | stretching of aliphatic -C-H | | |
| 2852 | methylene ($-CH_2-$) stretching | | |
| 2666[b]<br>2547[b] | combinational vibrational modes or overtones | | |
| | | 1724 | C=O stretching |
| 1715[c] | C=O stretching of esters | | |
| 1707[a]<br>1690[b] | C=O stretching of carboxylic acid and/or carbonyl groups of the photolysis products inside the chains | | |
| 1611 | aromatic ring C=C stretching | 1610 | aromatic ring C=C stretching |
| 1577 | C=O stretching of esters | | |
| 1504 | aromatic ring C=C stretching | | |
| sh1437+1450+sh1465<br>1425[b] | bending of $-CH_2-$ | 1464 | glycol C-H deformation |
| 1408<br>1391<br>1370<br>1338[c] | C-C-O, C-O-C, and C-C-H vibrations | | |
| | | 1278 | mixed mode of ring-carbonyl stretching, O-C stretching, and ring CH in-plane bending |
| 1241+sh1262 | C-O stretching of esters and ethers[d] | | |
| | | 1172 | ring CH in-plane bending |
| | | 1104 | mixed mode of ring CH in-plane bending, glycol C-O stretching, COC and CCO bending and C-C stretching |
| 1095+sh1116 | C-C-O and C-O-C of esters and ethers[d] | | |
| 1016 | C-O of glycol | | |
| 872 | C-H out-of-plain bending for para-substituted benzenoids (terephthalic acid) | | |
| | | 856+sh800/784 | ring C-C breathing |
| 727 | C-H deformation | | |
| | | 700 | ring C-C-C out-of-plane bending |
| | | 634 | ring C-C-C in-plane bending |
| | | 264 | mixed mode of ring C-C-C in-plane, and ring C-C out-of-plane |

Note: [a,b]observed only in spectra after UV exposure for 3872 and 7744h, resp.; [c]clearly distinguished only in original PET sample; [d]ethers such as diethyleneglycol and triethyleneglycol may serve as additives in PET.

For PLA, IR and RS spectra revealed no significant changes in vibrational pattern (neither in peak positions, nor in relative intensity of peaks within a particular spectrum) of the polymer during long-term UV exposures (see Figures 11a and 11b). Our IR spectrum of the original and UV-treated PLA corresponds well with the IR spectra of pristine PLA previously published in the literature [32, 35, 37, 43]. Peak positions and their assignment for PLA are clearly listed in Table 5.

Similarly as in Mikes et al. [43] who dealt with simultaneous exposure of 3D printed PLA plates to ozone and UV, we observed no influence of UV irradiation on IR and RS spectral pattern of PLA samples. On the contrary, UV impact on PLA structure was observed and discussed several times in the literature [32, 35, 37]. However, these authors [32, 35, 37] were applying not solely UV irradiation, but simultaneously increased temperature and relative humidity, which could lead to an easier photo-hydrolysis of PLA (manifesting itself in vibrational spectra by a decrease of C=O stretching and increase of -OH and COO- stretching vibrations). It should be also noted that they were applying UV irradiation of lower wavelengths than in our case, e.g., 315 nm in ref. [32] and 254 nm in ref. [35]. Moreover, both research groups were testing a thin film of PLA, which was measured in the transmission mode of IR absorption; whereas a 3D printed bulk cube is investigated in our study by the ATR mode (i.e., vibrational information is gained from near the surface). The structural stability of PLA with sole UV exposure is then responsible for the stable mechanical properties (see Fig. 6, Fig. 7, Fig. 8 and Fig. 9) and imparts PLA a good mechanical strength.

Concerning ABS vibrational spectra recorded before and after long-term UV irradiation (see Fig. 12), the situation is different than in PLA case. Distinct changes were observed in several parts of IR spectra of ABS when comparing UV irradiated samples with the original ABS spectrum (see Fig. 12a) and detailed IR spectra of ABS in Fig. SI8a-c; band assignment in Table 6): (i) increase in 3100 - 3600 $cm^{-1}$ region (characteristic O-H stretching vibrations), (ii) increase and broadening of 1727 $cm^{-1}$ band (characteristic C=O stretching), and (iii) decrease in 965 $cm^{-1}$ (C=C-H deformation of polybutadiene part). Furthermore, subtle changes were observed in RS spectrum of ABS after UV irradiation (particularly, a decrease of 1668 $cm^{-1}$ band and in the region around 1300 $cm^{-1}$, both being attributable to butadiene part of ABS – see Table 6 and Fig. SI8d). These nuances may not be observable directly in wide spectrum presented in Fig. 12b but can be clearly distinguished in more detailed spectra of ABS in SI (Fig. SI8d). Moreover, there is an increase of fluorescent background in UV-exposed ABS samples (Fig. SI8e). All the above-mentioned changes in our vibrational spectra report on photo-oxidative damage, crosslinking and chain scission (photo-degradation) of ABS.

The same changes in IR spectra of ABS exposed to natural and accelerated weathering have been reported by other authors [33, 36, 103]. The ABS investigated here most probably contains an additive or impurity, since the peak around 1727 cm-1 does not belong to the characteristic IR signal of ABS – there are no C=O groups in the structures of acrylonitrile, butadiene or styrene. This is supported independently by the Raman data: as in Mikes et al. [43], who used ABS with colouring additives, we observed changes in the fluorescent background of our ABS RS spectra. A carbonyl index (the absorbance of C=O divided by that of -CH2-) is commonly used to evaluate the weathering impact on polymers [103]; in our case it changed significantly with increasing UV exposure time, in agreement with the observed IR changes (detailed calculation in SI, part 10.4.1).

Many significant differences in IR spectra of non-treated and UV-treated PET-G samples were observed (Fig. 13a, band assignment in Table 7, detailed spectra in Fig. SI9a-c). Particularly, in the region of characteristic vibrations the following changes can be clearly distinguished in IR spectra of PET-G after long-term UV exposure (Fig. 13a): (i) enormous increase of O-H stretching vibrations (3180 – 3530 $cm^{-1}$); (ii) changes in relative intensity of aromatic vs. aliphatic CH stretching vibrations (above and below 3000 $cm^{-1}$, respectively); and (iii) appearance of overtones and/or combinational vibrational modes (2550 – 2650 $cm^{-1}$) – for details see Figs SI9a-c. In the fingerprint region of IR spectra in Fig. 13a and Fig. SI9b, broadening and shift of the position of absorption maximum of C=O vibration (1690 – 1715 $cm^{-1}$) dominates among the other changes. Indeed, the maximum of carbonyl band moves from 1715 $cm^{-1}$ (in original PET-G) through 1707 $cm^{-1}$ (in 3872h UV irradiated sample) to 1690 $cm^{-1}$ (in 7744h UV irradiated sample). This can be related either to changes in chemical character of C=O bond in polymer chains, from esters (in original PET-G) to free carboxylic acids -COOH (in photo-degraded PET-G), or to occurrence of new carbonyl bonds being present in products of photolysis of polymeric chains due to UV irradiation. Moreover, a new peak around 1425 $cm^{-1}$ evolves while the peak at 1338 $cm^{-1}$ disappears, which could be related to changes in $-CH_2-$ bending and in C-C-O, C-O-C, C-C-H vibrations (Table 7). The overall broadening of the original PET-G IR absorption bands in the region below 1300 $cm^{-1}$ in IR spectra of UV-treated samples (Fig. 13a and Fig. SI9c) is most probably caused by the presence of many other types of deformation and stretching vibrational modes being attributed to photoproducts. Hence, the final IR spectral pattern represents an envelope of all species' vibrational contributions.

There are distinct differences in baseline-corrected RS spectra of the original PET-G sample and of UV-treated ones normalized to the maximal signal within each spectrum (Fig. 13b). The as-measured RS spectra of PET-G are highly masked by fluorescent background as can be evidenced in Fig. SI9d-e. Importantly, the fluorescent background is the highest for 3872h-UV treated sample of PET-G and substantially decreased in the case of 7744h-UV treated sample. It should be kept in mind that RS spectra for PET-G were measured with a reduced laser power (30 mW) to avoid sample degradation and burning (as a black sample, PET-G in our study is absorbing even the employed 785 nm excitation laser line); therefore, their signal-to-noise ratio (SNR) is much worse than in PLA and/or ABS cases (compare Fig. 11b and Fig. 12b with Fig. 13b). Nevertheless, in the original sample (UV untreated), several characteristic peaks of PET can be distinguished, such as 1724, 1610, 1464, 1278, 1104, 856, 700, 634, and 264 $cm^{-1}$ (for peak assignments see Table 7). Owing to the strong fluorescent background, it is possible to find only three RS spectral features of PET in UV-irradiated samples whose positions seem to remain unchanged (Fig. SI9f): C=O (1724 $cm^{-1}$), aromatic ring C=C stretching (1610 $cm^{-1}$), and mixed mode of ring-carbonyl stretching, O-C stretching, and/or ring CH in-plane bending (all of them being located at around 1300 $cm^{-1}$). Due to the peculiarities with fluorescent background, consequent automatic baseline correction, and poor SNR, the quantification of the changes potentially made from RS spectra is not possible.

All the above-mentioned changes in vibrational spectra of UV-exposed PET-G in comparison to original PET-G have been previously described in the literature [31, 34, 38, 39, 94, 101, 103, 104] and are attributed to photo-oxidation damage, polymer chain scission, which appears namely in the ester bond -(CO)O-$CH_2$-. Consequently, it leads to molecular weight decrease, changes in morphology and mechanical properties of PET. On the contrary to PLA and/or ABS cases, the researchers dealing with PET weathering agree on its degradation (one of the main products of photolysis is terephthalic acid [101]) albeit they are using different conditions of weathering, dissimilar types of PET films of various thicknesses, and many types of UV radiation sources. For instance, similar broadening of characteristic carbonyl band in IR spectrum of PET and simultaneous augmentation of characteristic absorbance of O-H vibrations as in our case was observed in [103]. The authors [103] noted that it was difficult to assess the oxidation degree of PET based on the formation of carbonyl species. On the other hand, Pires et al. [38] exploited characteristic C=O vibration of esters (being characteristic for the polymeric backbone of PET only) and estimated the carbonyl index in a different manner: as the ratio of absorbance at 1720 $cm^{-1}$ vs. absorbance at 1508 $cm^{-1}$ (i.e., carbonyl of esters vs. aromatic C=C stretching which represent an unalterable band). The resulting carbonyl index in their case [38] slightly decreased for PET after the exposition to UV which means that polymeric chain scission occurred (in other words, the relative content of ester bonds decreased). It should be noted here that there is no exact definition of the carbonyl index as already stated in [103]. Therefore, it might be puzzling that in ABS case the degradation of polymer is expressed as the carbonyl index increase, while in PET case as its decrease. In general, the carbonyl index is a measure of changes in characteristic bonds in a particular polymer, reporting thus about their degradation degree regardless of whether the values show a positive or a negative trend.

Inspired by the latter approach [38], we applied this type of carbonyl index calculation for our PET-G case (i.e., using the values of absorbance at 1715 $cm^{-1}$ and at 1504 $cm^{-1}$, respectively) - see Table SI2. The carbonyl index of PET-G after long-term UV exposures decreased in our case substantially in comparison to the original PET-G: from the value of approx. 11 to approx. 4. Interestingly, according to the calculated carbonyl index of PET-G, the extent of PET-G degradation induced by UV in the sample exposed for 3872h is practically the same as in the sample irradiated for 7744h. The negligible difference in the carbonyl index of both UV-treated samples of PET-G may be caused by the fact that the degradation of the polymer reached a plateau after certain period of UV irradiation. Two main factors may induce such a plateau formation: (a) penetration depth of UV radiation affecting the sample properties can be reached in a certain distance from the surface and does not propagate any deeper into the sample (i.e., optical reason). (b) Measurable depth (that is constant in our ATR FT-IR setup for a given type of material) does not enable to investigate deeper layers of PET-G samples (i.e., instrumental reason). Nevertheless, the carbonyl index determined for PET-G corroborates quantitatively the pronounced photo-degradation of our PET-G exposed to long term UV irradiation.

The significant differences (while possessing the same carbonyl index) in IR spectral patterns of 3872h- and 7744h-UV exposed PET-G samples (Fig. 13a) can be ascribed to the presence of different concentrations and types of photo-degradation products such as fluorescent chromophores at PET-G surface. Direct evidence about their presence in untreated and UV-treated PET-G samples was given by our RS spectra as well. Studying literature, we revealed that fluorescence of PET is a known fact (e.g. [31, 34, 42]). Indeed, UV irradiation of PET leads to the formation of mono- and di-hydroxy terephthalates which are manifesting themselves by fluorescence in the region of 350-550 nm [31, 34]. Fluorescence of PET is increasing in intensity when UV is applied, may be photobleached and after prolonged UV-exposures, it is bathochromically shifted [34]. This well explains why a much higher fluorescent background was observed in our RS spectra of 3872h-UV exposed PET-G sample than in that of 7744h-UV exposed one. Finally, PET-G contains glycol additives, as noted in the Experimental section. Diethylene and triethylene glycol act as end-capping agents, but in larger amounts they also provide preferential sites for hydrolytic attack and contribute to the formation of hydroxyterephthalate chromophores [94].

In summary, IR and RS spectra revealed no significant changes in PLA even after 7744 h, whereas ABS and PET-G showed pronounced changes in their vibrational patterns, quantified by the carbonyl index and most pronounced for PET-G. These results are consistent with the mechanical data.

### 3.4 Control samples stability

Polymers can also change with time at room temperature in the absence of any external agent, and such a contribution could act as a secondary factor in the changes observed after UV exposure. A set of samples from the original batches was therefore retained as a control, kept wrapped in a closed cabinet in a temperature-stabilised air-conditioned laboratory for seven years after the original measurement, and re-measured using the same indenter tip, measurement protocol and calibration procedure, with a reproducibility of ±4 %. An incomplete intermediate set measured after approximately four years is also included. Fig. 14 shows hardness *H* and reduced modulus *E*r at 2 mN, chosen because at 0.5 mN the results are strongly affected by surface inhomogeneity while 2 mN still samples the near-surface region.

Six of the eight materials show an increase in both quantities; for five of them it is modest, between +1 and +19 % in *H* and +4 and +13 % in *E*r, the best-defined being HIPS (white) (+16.5 % and +12.8 %, with the lowest scatter of the set), while only ABS (white) is unchanged within the reproducibility of the measurement. PC (white) shows a larger apparent increase (+48 % in *H*), but this material has by far the largest scatter, and the change falls close to its own standard deviation. An increase in hardness and modulus during storage below the glass transition temperature is the expected consequence of physical ageing, the spontaneous structural relaxation of a glassy polymer, which reduces free volume and increases density and hardness [105, 106]; since the intermediate point does not show a monotonic progression, physical aging is suggested here as a possible factor rather than as a proven mechanism. Two materials behave differently: PLA (white) decreases progressively (-21.9 % in *H*), whereas PLA (black), which differs only in pigmentation, increases slightly, and the reason is not resolved by these data; ABS-T (black) also decreases (-42.8 %), but with a very high relative standard deviation of 41.5 %, and also higher scatter in lower load (0.5 mN, not shown here) this control measurement is therefore considered inconclusive.

At 2 mN, UV exposure for 7744 h increased the hardness of ABS (white), ABS (black), PETG (black) and HIPS (white) by +89 to +195 %, that is five to more than a hundred times the change measured on the corresponding stored controls, so that for these materials the effect of UV exposure is undeniable. For PC (white) the UV-induced increase (+69 %) is of the same order as the control drift (+48 %) and the two contributions cannot be separated with confidence. For both PLA grades the changes after UV exposure (+8 to +13 %) are comparable in magnitude to those of the controls and are not regarded as significant. It should be noted that the control data quantify the drift accumulated over seven years, whereas the UV exposure and the corresponding measurements were completed within approximately one year of the original unexposed measurement. Since structural relaxation in glassy polymers proceeds approximately logarithmically with time, the ageing contribution present in the UV data is smaller than the seven-year values, which therefore represent an upper bound on the storage-related component of the reported UV effects [105].

The control specimens were also examined by laser scanning confocal microscopy. They show only the texture formed against the build plate and none of the crack networks found after UV exposure (Section 3.3.1). Regardless of its role in the aging process, storage does not cause any surface cracks, and morphological damage can therefore be attributed solely to the effects of UV radiation.

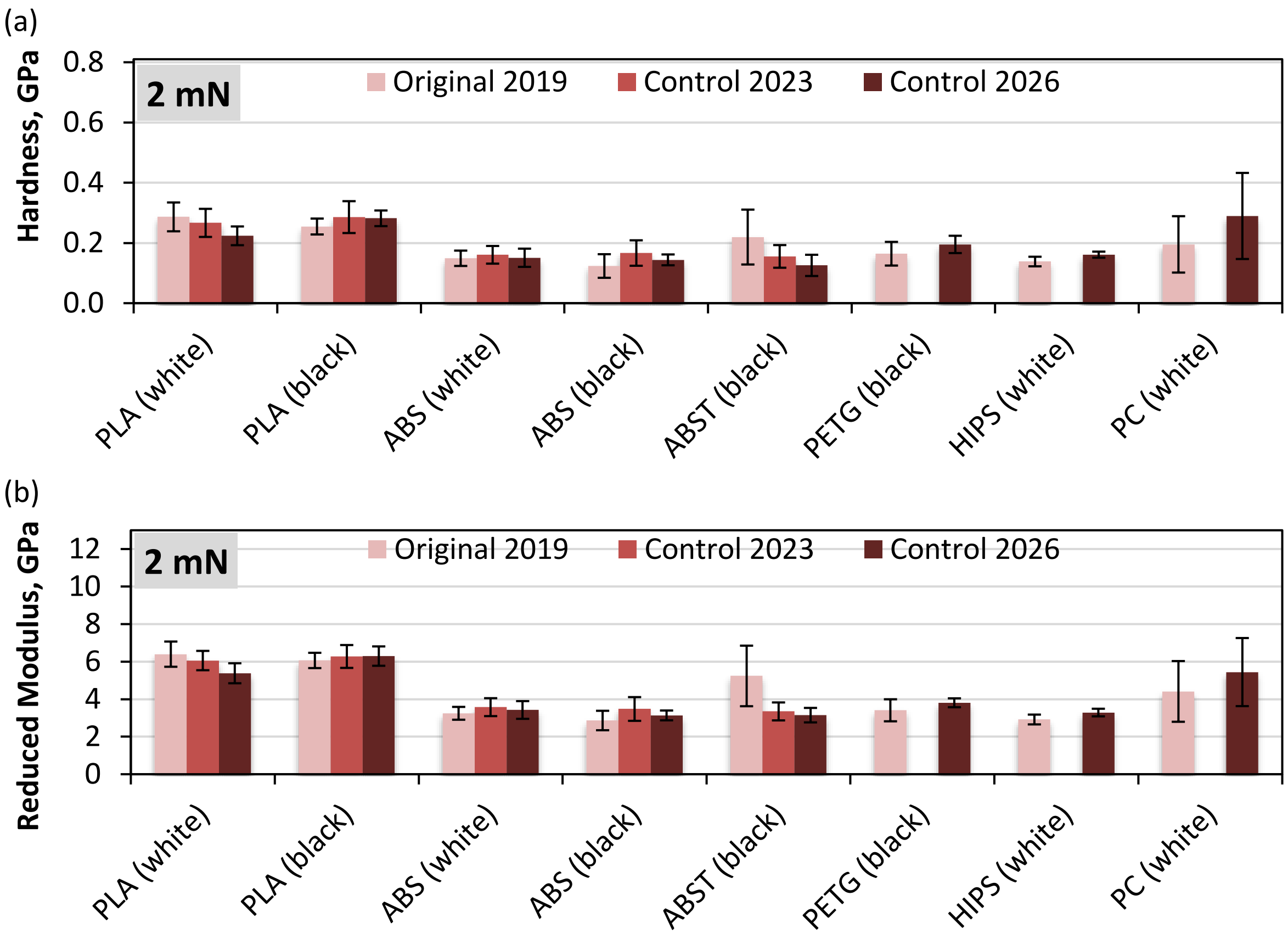

Fig. 14. (a) Nanoindentation hardness $H_{IT}$ and (b) Reduced modulus $Er$ for polymer samples comparing the original unexposed set against two (2023, 2026) re-measurement of control set.

## 4 Summary

The selected polymeric samples PLA, ABS, ABST, PETG, HIPS and PC, representing the most widely used polymeric materials for 3D printing in practice, were produced by 3D-print based FFF method. The resistance of the samples to photodegradation was studied, induced by continuous long-term UV exposure in two steps - 3872 hours and 7744 hours, i.e. up to 322 days. Changes in mechanical properties of samples (hardness, elastic modulus, creep) were monitored using local sensitive method of nanoindentation; whereas qualitative and quantitative (carbonyl index calculations) structural changes induced by UV long-term exposure within selected representative samples were evaluated using vibrational spectroscopic methods (Infrared-IR absorption, Raman spectroscopy-RS).

The samples demonstrated varying degrees of resistance to photodegradation. The PLA samples did not undergo any observable changes in either mechanical, or structural properties. In contrast, ABS and ABST samples showed an increase in surface hardness, consistent with reduced molecular mobility and increasing brittleness of the surface layer; this interpretation is supported by the surface cracks observed by confocal microscopy and by significant changes in vibrational spectra indicating crosslinking and chain scission (photo-degradation). The most significant changes in local mechanical properties on the sample surface were found for PETG samples. This was confirmed by qualitative and quantitative results from spectroscopic methods. Moreover, the advanced indentation approach, reaching a larger depth, revealed the limitation of photo-induced changes to surface regions in PETG samples in comparison to ABS(T) or HIPS samples. Furthermore, HIPS, often used only as a stabilizing support for printing or as the cheapest option for the final product, also underwent photodegradation as expected.

The PC sample showed an increase in near-surface hardness and modulus of approximately 80 % after 3872 h at the lowest indentation load, with little further change thereafter. This increase is, however, accompanied by the largest scatter of the whole set, and is of the same order as the drift measured on the stored control specimens over a comparable period (Section 3.4), so the photo-oxidative and storage contributions cannot be separated with confidence for this material. Creep measurements in the near-surface region indicated a marked decrease of creep mainly in ABS(T) and HIPS samples with length of UV exposure, whereas PETG showed only a small decrease, PC practically negligible and PLA no effect. The most significant change in creep occurred between the original sample and the 3872-hour exposure, while the doubling of exposure to 7744 hours resulted in only minor change.

The crack networks observed by confocal microscopy provide evidence of embrittlement that is independent of the indentation measurements. Whereas the increase in hardness and reduced modulus describes the deformation response, the appearance of surface cracking is a change in fracture behaviour and therefore addresses embrittlement directly. The two observations are mutually consistent: the materials that developed crack networks (ABS, PETG, HIPS, PC) are those for which the near-surface hardness increased, while PLA remained unaffected, showing neither cracking nor a significant change in hardness.

The presented results establish how UV radiation affects the near-surface mechanical properties of the tested materials, isolating and quantifying a single weathering agent. They apply to the specific commercial formulations investigated, whose additive packages – in particular the type and concentration of UV absorbers, light stabilisers and antioxidants – were not disclosed by the manufacturers and are therefore unknown. The ranking obtained should accordingly not be generalised to all materials belonging to the PLA, ABS, PETG, HIPS and PC families. The exposure represents isolated UV irradiation rather than complete outdoor weathering, and the results cannot be regarded as a substitute for a full weathering assessment.

## 5 Acknowledgement

This research was funded by Operational Programme Research, Development and Education project No. CZ.02.1.01/0.0/0.0/17_049/0008422 of the Ministry of Education, Youth and Sports of the Czech Republic. Authors also acknowledge Dr. Jakub Navarik (Palacky University, Olomouc, Czech Republic) and companies Prusa 3D print (Prague, Czech Republic) and Filament Plasty Mladeč (Haňovice, Czech Republic) for production and supply of samples. Authors also acknowledge Dr Martin Vacula for performing the photometric measurements on the lamp. DeepL translator and Claude.ai was used to refine English grammar and phrasing of some part of the.

## 6 CRediT authorship contribution statement

Jan Tomastik oversaw supervision, conceptualization and design of the study and methodology; performed and evaluated majority nanoindentation experiments; wrote the original draft of the manuscript. Karolina Siskova designed and performed IR and RS measurements, treated and evaluated vibrational spectra; edited and corrected the original draft of the manuscript. Both were in charge of funding acquisition. Radim Ctvrtlik performed and evaluated nanoindentation experiments; contributed to the overall design of the manuscript. Lukas Vaclavek performed and evaluated creep experiments. All the authors have contributed equally on drafting and revisions of the manuscript.

## 7 Declaration of competing interest

The authors declare that none of them has any known competing financial interests or personal relationships that could influence the research described in this article.

## 8 Appendix - Supplementary materials

The Supporting Information section includes an extension of the Experimental section—a description of nanoindentation methods and examples of load-displacement curves. It also provides a detailed explanation of the method for evaluating creep from nanoindentation tests using three creep parameters. This section also expands on the Results and their discussion—nanoindentation results for a wider range of samples, which were reduced in the main article (minor changes); a comparison of the same materials from different manufacturers; and an extension of data from the LPU method and creep measurements for UV exposure over a shorter period of 200 hours. The analysis of structural changes is supplemented with details from the IR and RS spectra of selected samples and detailed data on the carbonyl index.

## 10 Supporting information

### 10.1 Samples

In addition to selected samples from ABS and PLA, other samples with additives were also supplied as part of the initial sample selection. However, in both cases of **PLA with steel powder** and **ABS with graphite powder** samples could not be measured satisfactorily due to the large scatter in the indentation data implicating higher roughness and inhomogeneity. Therefore, the samples were ultimately removed from the research set and not further analysed. However, motivation for further research may focus on such doped samples, which are an obvious trend in the 3D printing community.

### 10.2 Nanoindentation testing

The principle of the nanoindentation method consists in the penetration of a diamond tip of known geometry into the sample surface with precisely defined force. During the test, the load and depth are continuously recorded. Subsequently, the load-displacement curve is evaluated by standard method [57] to obtain information on mechanical properties such as hardness, modulus of elasticity, elastic/plastic work and creep parameters. There are several variants of indentation test based on loading-unloading schedule.

Two main variants employed in this research are standard nanoindentation with several (10 — 15) indents to several places to obtain statistics about certain area. On the other hand, load-partial unload method (LPU) assesses mechanical parameters on one spot only, but with several load-unload cycles obtains mechanical property values from different depths. Indentation curves (dept-load) of both variants are shown for tested samples Fig. SI1.

(a)

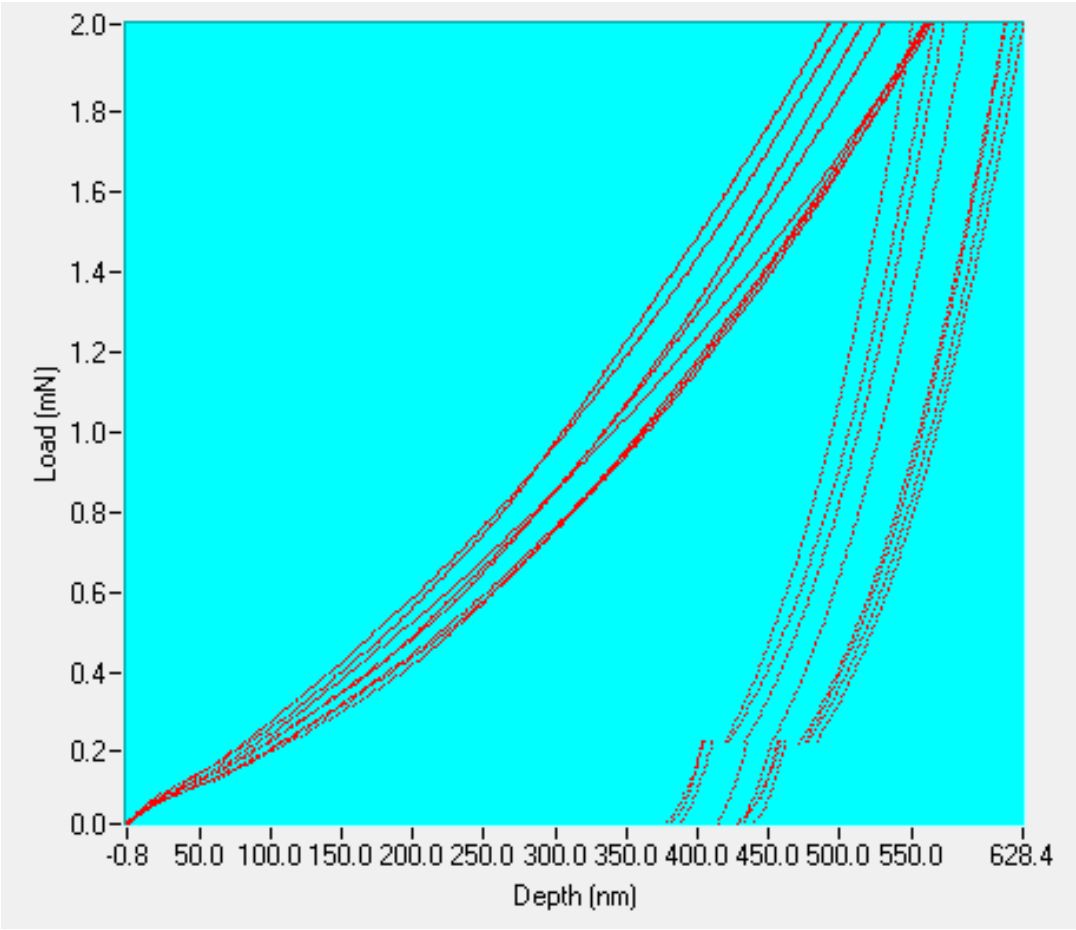

(b)

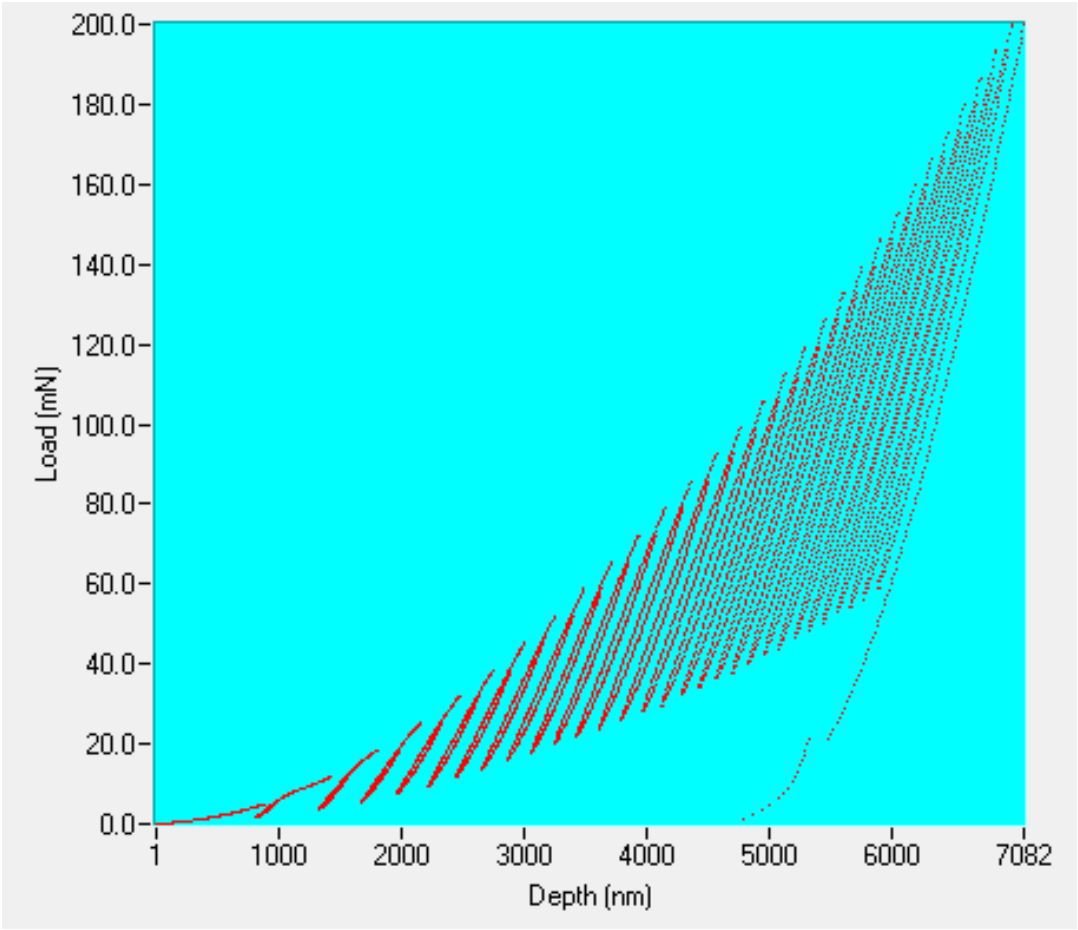

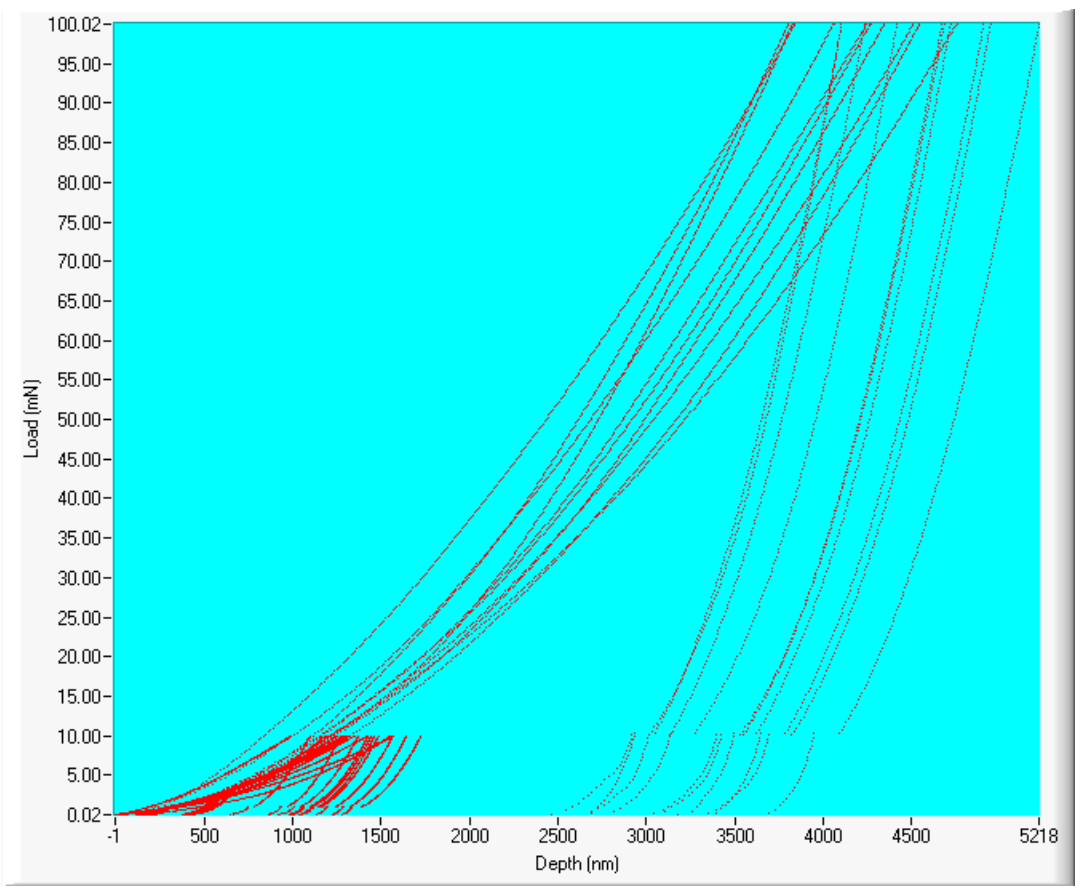

(c)

Fig. SI1. (a) Standard nanoindentation curves of load-depth record. Spread of several measurements in one sample. (b) Load-partial unload (LPU) method and resulting load-depth record. (c) Comparison of nanoindentation curves for 10 mN (smaller) and 100 mN (larger).

### 10.2.1 Nanoindentation results

More detailed indentation results with three more samples added to the set i) PLA-white-b ii) PETG-black-b, iii) PETG-black-c, are shown in Fig. SI2 for hardness and in Fig. SI3 for elastic modulus. Also graph for load of 100 mN is added up to the 0.5mN, 2 mN and 10 mN from the article´s Fig. 6 and Fig. 7.

(a)

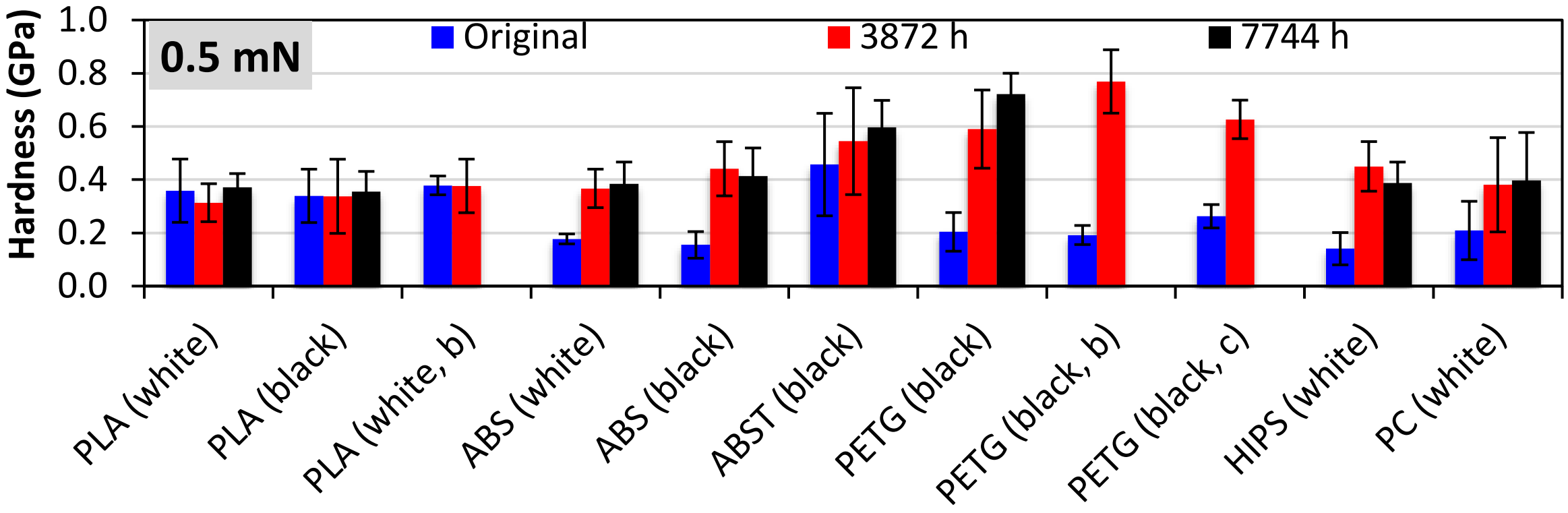

(b)

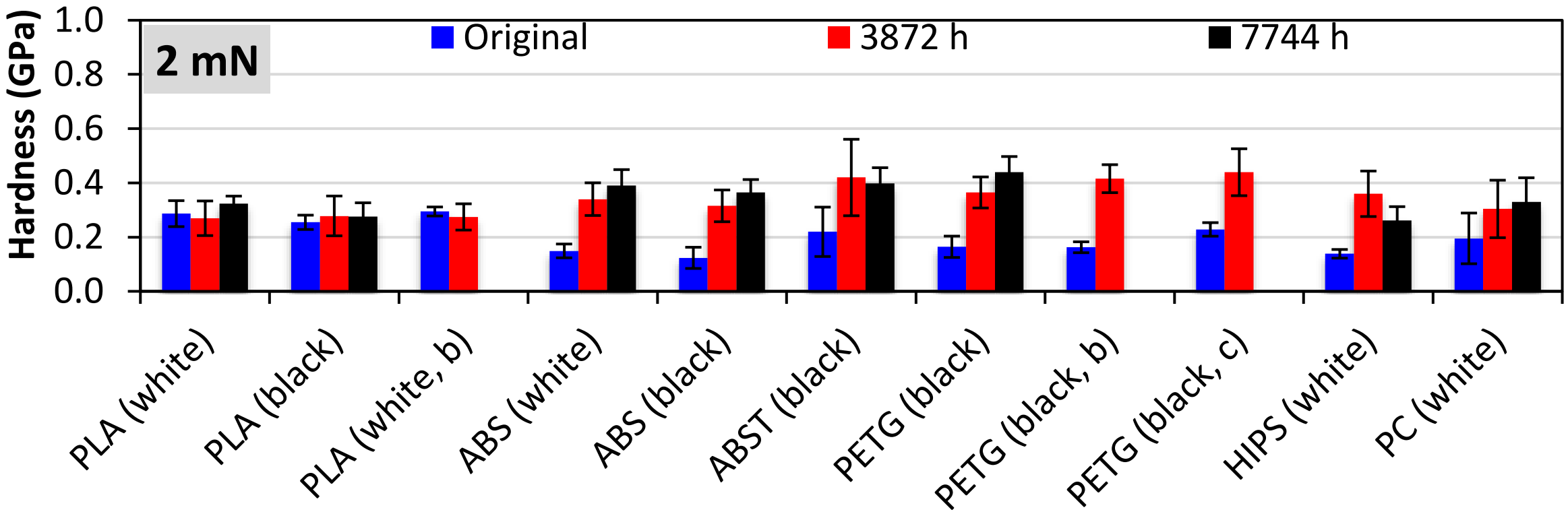

(c)

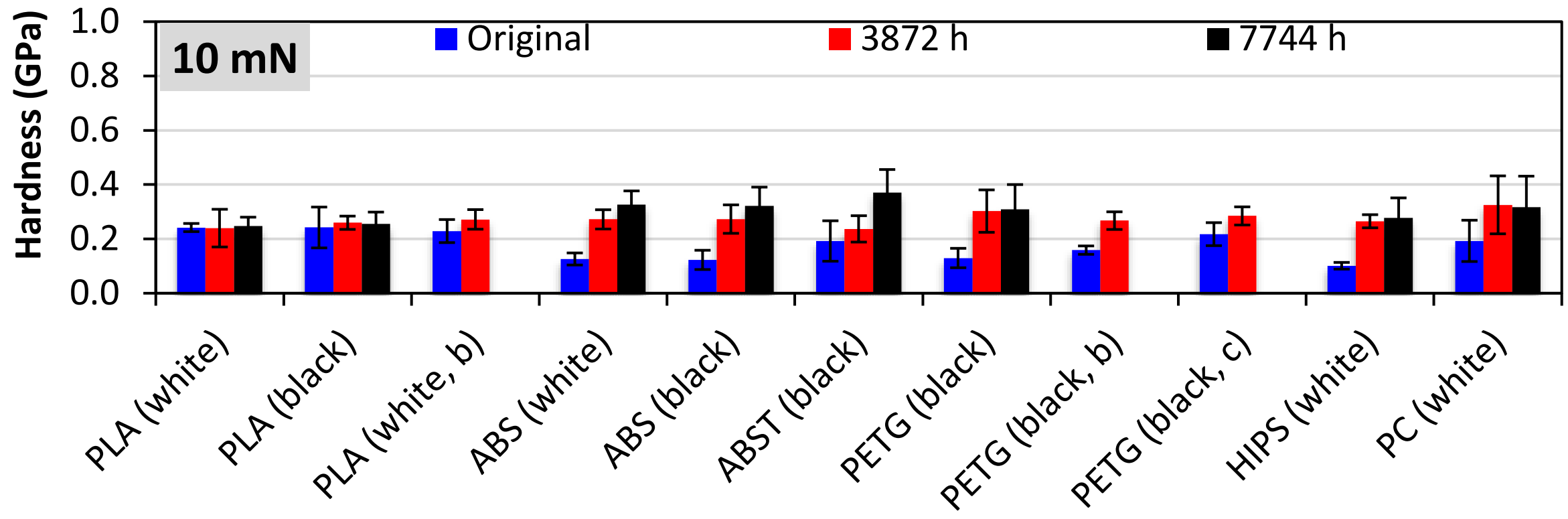

(d)

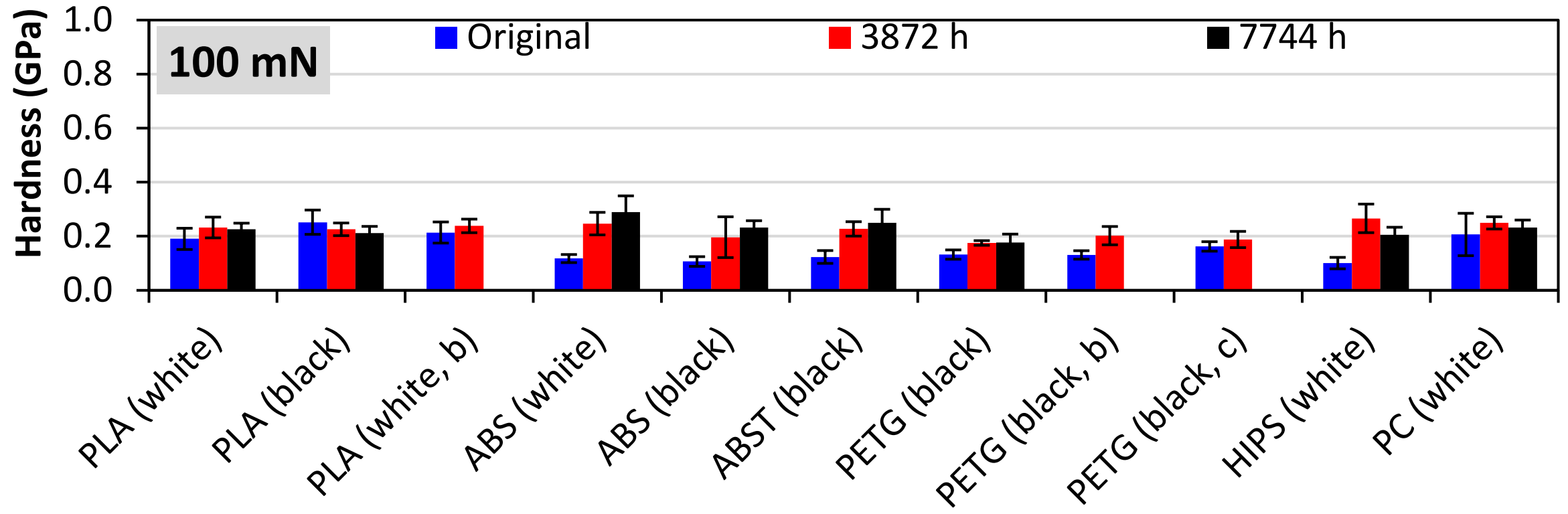

Fig. SI2. Nanoindentation hardness $H_{IT}$ for selection of samples from Table 1. comparing values for original state, 3872 hours and 7744 hours of summarized UV exposure obtained for load of (a) 0.5 mN, (b) 2 mN, (c) 10 mN and (d) 100 mN.

(a)

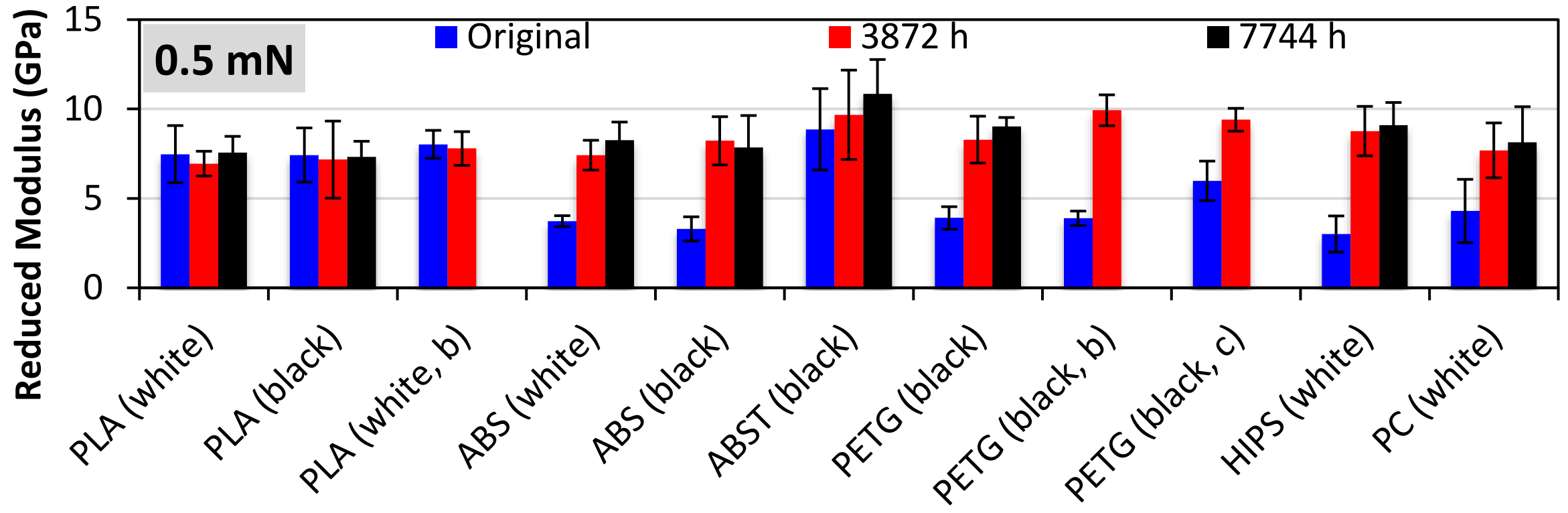

(

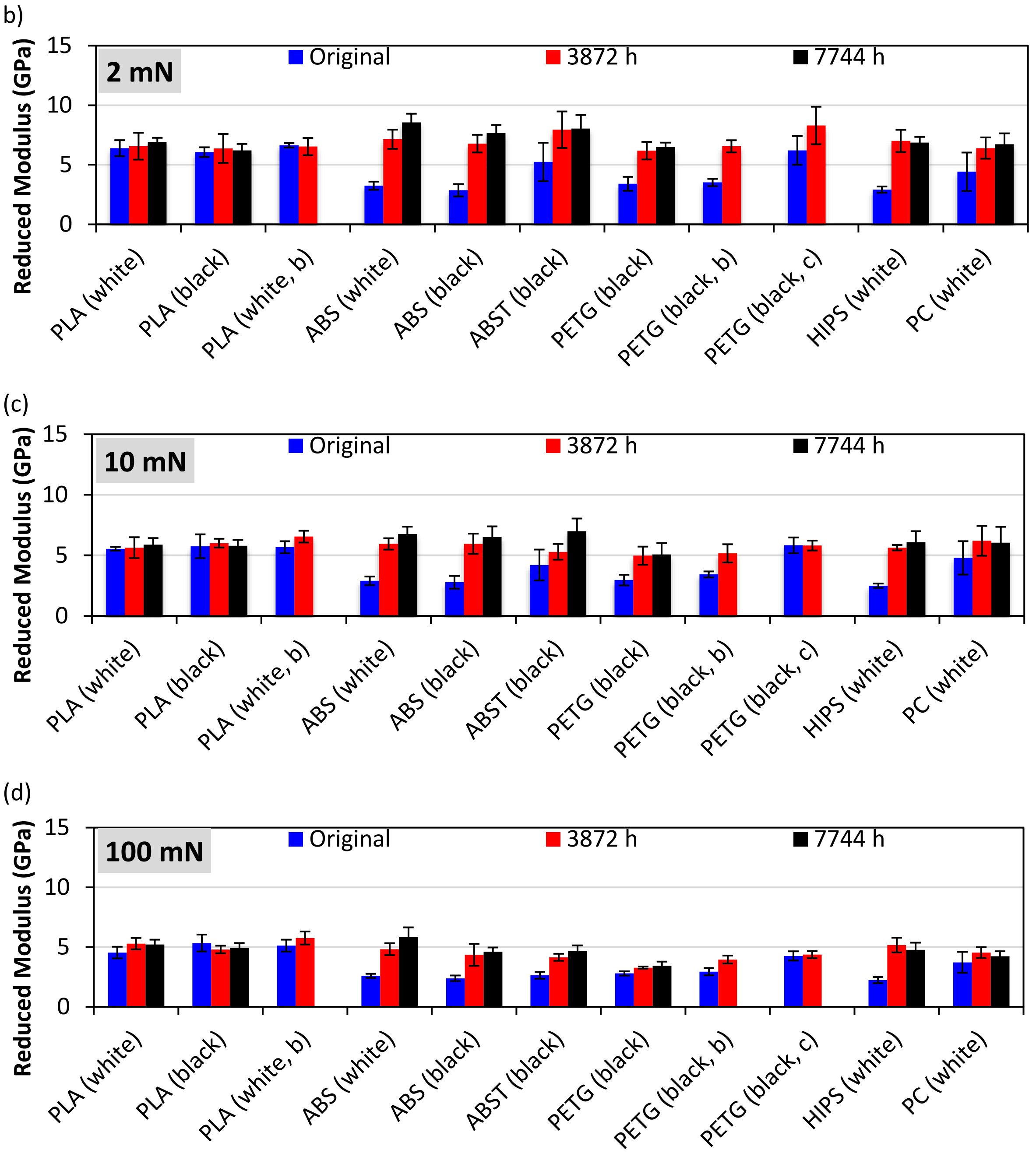

Fig. SI3. Nanoindentation reduced elastic modulus $E_r$ for selection of samples from Table 1 comparing values for original state, 3872 hours and 7744 hours of summarized UV exposure obtained for load of (a) 0.5 mN, (b) 2 mN, (c) 10 mN and (d) 100 mN.

In addition to the original three PLA samples (PLA black, PLA white, PLA white-b) two other PLA samples (PLA white-b+ and PLA white-b++) from same manufacturer (Prusa 3D) as PLA white-b but with UV-stabilizer additive of volume content 1 and 3 %. The results of hardness were identical with no difference observed after 3872 h of UV exposure (see Fig. SI4). PLA white-

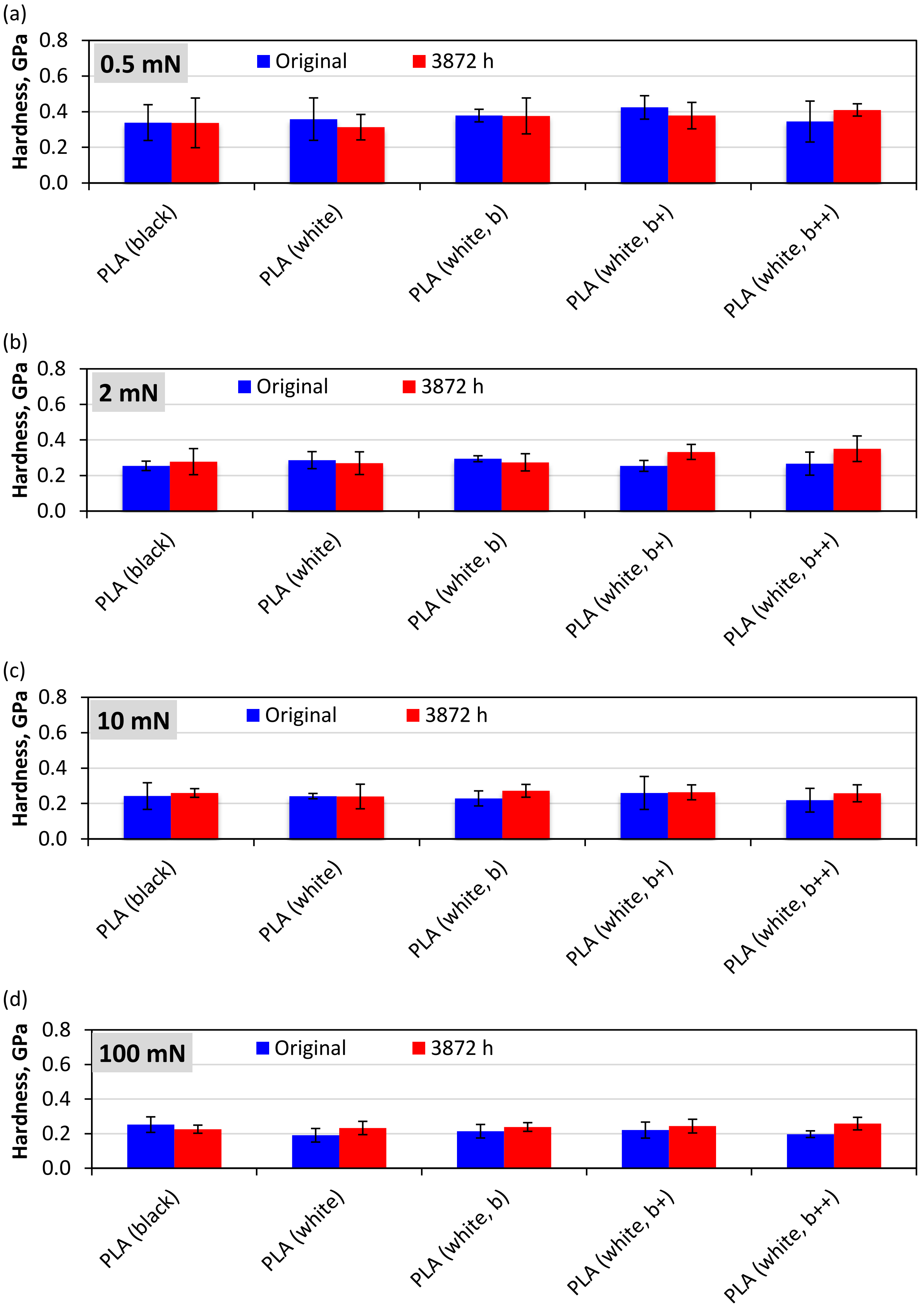

Fig. SI4. Nanoindentation hardness $H_{IT}$ for PLA samples from different manufacturers comparing values for original state and 3872 hours of summarized UV exposure obtained for load of (a) 0.5 mN, (b) 2 mN, (c) 10 mN and (d) 100 mN.

### 10.2.2 Load-partial unload results

Alternative load-partial unload tests were also applied next to the standard indentation. Results were discussed in the main article (see Fig. 8). In addition to the data from the measurements of the unaffected (Original) and long-term UV affected (3872 h and 7744 h), other set of data from 200 hours UV exposure was added. This dataset sometimes shows gradual effect of UV exposure, but was measured with significant standard deviation, thus was omitted in main article. Here LPU test of ABS shows no apparent difference in comparison to original sample, however for HIPS already some effects are visible in lower depths, see Fig. SI5 for hardness and Fig. SI6 for the reduced modulus.

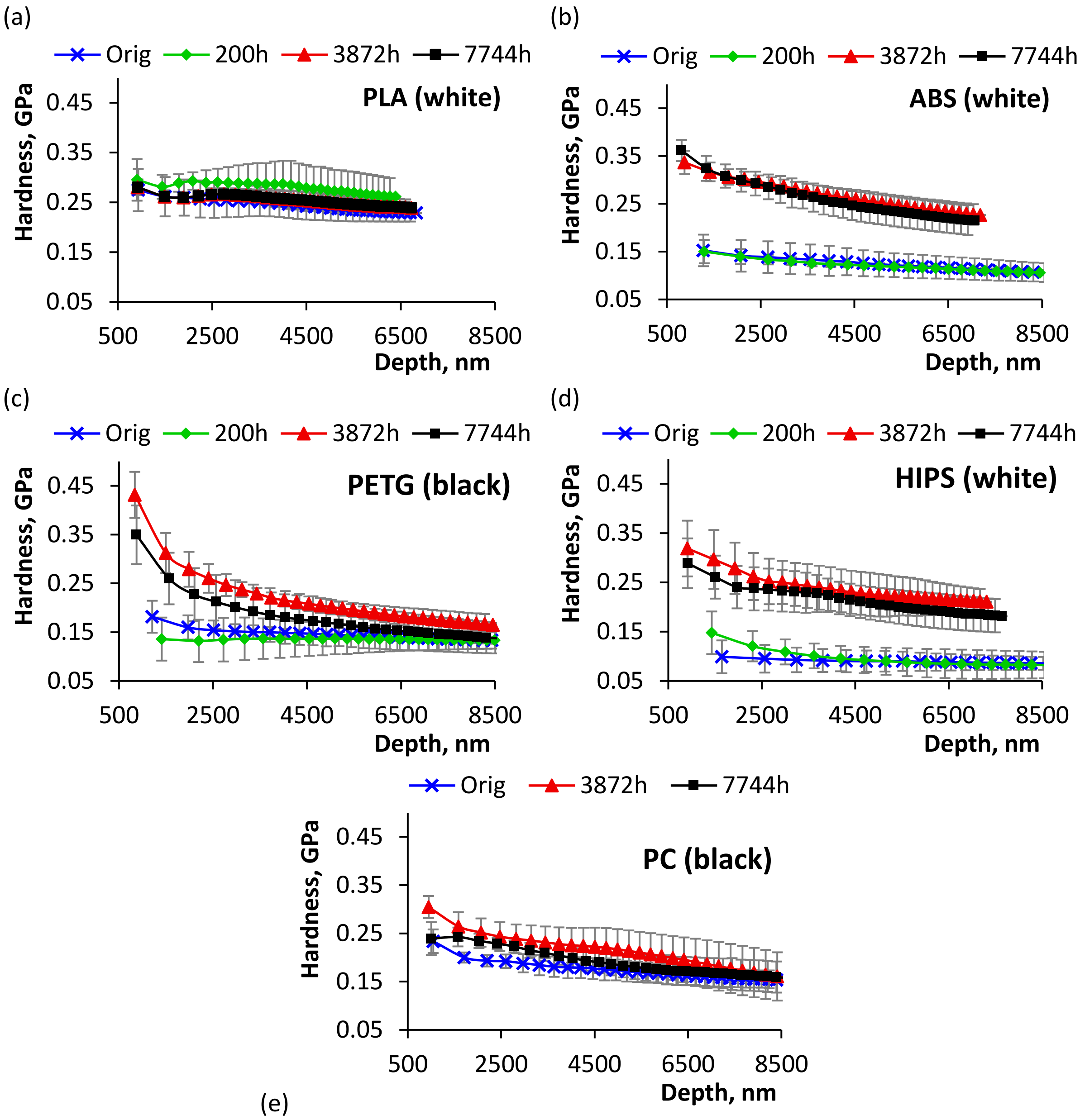

Fig. SI5. Depth profiles of nanoindentation hardness from load-partial unload (LPU) tests for (a) PLA w, (b) ABS w, (c) PETG b, (d) HIPS w and (e) PC w samples comparing original states and UV exposure after 200 h, 3872 h and 7744 h.

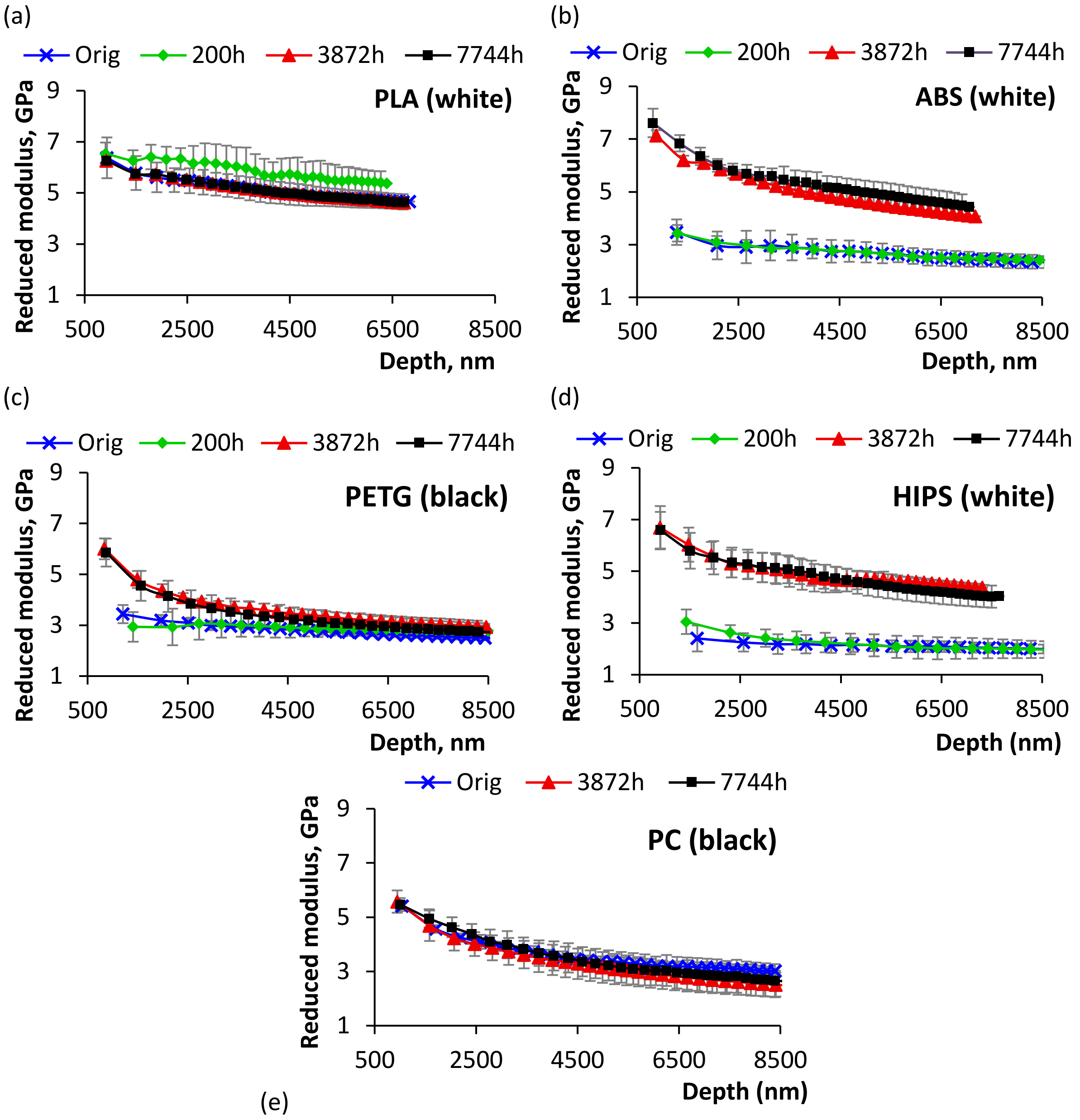

Fig. SI6. Depth profiles of nanoindentation reduced elastic modulus from load-partial unload (LPU) tests for (a) PLA w, (b) ABS w, (c) PETG b, (d) HIPS w and (e) PC w samples comparing original states and UV exposure after 200 h, 3872 h and 7744 h.

## 10.3 Creep - nanoindentation testing

We used nanoindentation tests with sufficiently long holding at maximum load to monitor creep of samples. However, several approaches are used in scientific practice for indentation creep analysis, which will be discussed further.

A transition from primary creep to secondary creep can occur on a relatively short or long-time scale, depending on the material. The former describes a significant deformation at the beginning of the test (first several or tens of seconds) while the latter describes a steady state deformation of smaller magnitude [107]. In standard tensile tests, creep is obtained on a scale of tens of minutes to hours. Nanoindentation methods are performed on a time scale of tens of minutes at most. However, several experiments using thermally stabilized depth-sensing indentation were successful in obtaining reliable indentation creep data with a noticeable transition from primary to secondary creep and hence allowed to obtain creep parameters [59, 108-110]. The standard method for evaluating indentation creep is described in the ISO standard [58], giving the **relative increase in depth during creep** period using equation

$$\text{relative creep [\%]} = (h_2 - h_1)/h_1 * 100 \tag{1}$$

where $h_1$ and $h_2$ are depths at the beginning and at the end of creep period, respectively (see Fig. SI7). Although the relative creep value gives information about the extent of creep over a specified selected time, it does not provide information about the creep development of the specimen in the future. In other words, it is not clear whether the material is still in the rapid phase of primary creep or whether it is already in the slower part of secondary creep or has even stopped creeping at all. The future evolution of creep is thus better described directly by the **slope of the dh/dt curve**, which can be obtained as a tangent to the final part of the curve as shown in Fig. SI7. However, this parameter has the opposite problem that it does not take into account the existing extent of creep, i.e. the intensity of primary creep. Goodal et al. [59] therefore proposed the novel **parameter $P_{CR}$**, accounting the both aspects – the range of the creep for the specified test conditions (temperature, indentation load and length of the hold period) and the slope of the final phase of the observed creep. $P_{CR}$ is the product of both components according to the equation

$$P_{CR} = (h_2 - h_1) * dh/dt \tag{2}$$

resulting in the number with units of $m^2/s$. If the $P_{CR}$ is low than the creep in material was and should have remain low. The high value indicates a tendency for the material to creep significantly in the future [60].

The analysis of the mentioned creep parameters and their informative value can be illustrated by the model curves in Fig. SI7. The creep curve "A" exhibits steep primary creep which quickly transitions to a phase of slower secondary creep characterized by a smaller value of the slope dh/dt. Curve "C" and, in particular "B", show a higher slope value. While curve C reaches only to half of depth increase through the given time period in comparison to curve A, it can be assumed that it would reach the same depth over a longer period of creep time. This is well reflected in the parameter $P_{CR}$, which would give comparable values for curves A and C. On the other hand, comparison of the A and B curves shows the same relative increase in creep depth over the defined time period, however the greater slope of the secondary creep curve in case of curve B leads to a prediction of clearly higher creep in the future. The $P_{CR}$ parameter for the B curve would therefore be the highest compared to the remaining curves. Thus, it can be said that the relative creep increment is suitable for comparative comparison of samples over a defined time, but the slope of the curve and the $P_{CR}$ parameter better predict creep evolution in future.

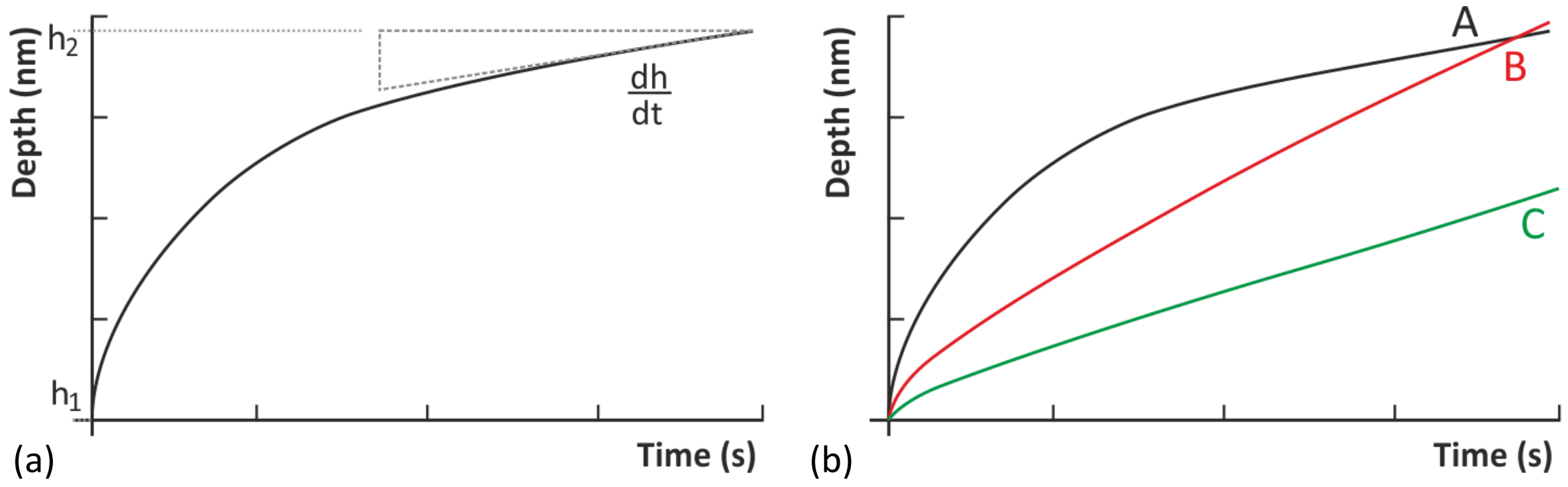

Fig. SI7. (a) Schematic of the indentation creep curve with parameters for its evaluation. (b) Model creep curves.

In this research, all three creep parameters were reported for examined samples. Tests were conducted using two values of maximum loads of 20 mN and 50 mN and an extended holding time of 300 seconds at the maximum load. However, results are presented only for the lower value of 20 mN, which is more surface sensitive. Ten measurements were always taken for each experimental setup; curves of non-standard shape were excluded from the analysis (in general 0 – 30 %). The slope of the curve was determined by tangent to the region of its last sixth, i.e. in the section between the 250th and 300th second of the creep measurement period.

## 10.4 Structural examination using IR and Raman spectroscopy

Structural changes in UV affected polymeric samples were studied using Raman scattering (RS) and Infrared absorption spectroscopy (IR). While IR absorption is often used in investigation of polymer structural changes induced by weathering [31-38, 94, 103, 104, 111] RS has been exploited less frequently [39-41]. The combination of IR and RS has been employed mostly in recent studies [42-47]. Generally, there are different selection rules for IR- and Raman-activity of vibrational modes of species: while IR absorption requires that molecules possess dipole moments, and their changes during vibrational motion are evaluated; RS is based on changes of polarizability tensor of molecules. In other words, and simplified, polar bonds can be investigated rather by IR absorption, whereas nonpolar bonds by RS. In this sense, combined information gained from IR and RS data is valuable. Moreover, RS is very sensitive to fluorescence, which represents a parasitic signal in the viewpoint of RS because it is more intensive. Nevertheless, changes in fluorescent background of RS spectra can give further information about samples.

The results of both spectroscopic methods were discussed in the main article comparing the original samples for the selected materials PLA, ABS, and PETG. The appropriate vibrational spectra for these polymers are shown in Figures 11-13, band assignments are listed in Tables 5-7. Detailed graphs of the IR and RS spectra, along with the carbonyl index, are provided here as Figures SI8, SI9 and Tables SI1 and SI2.

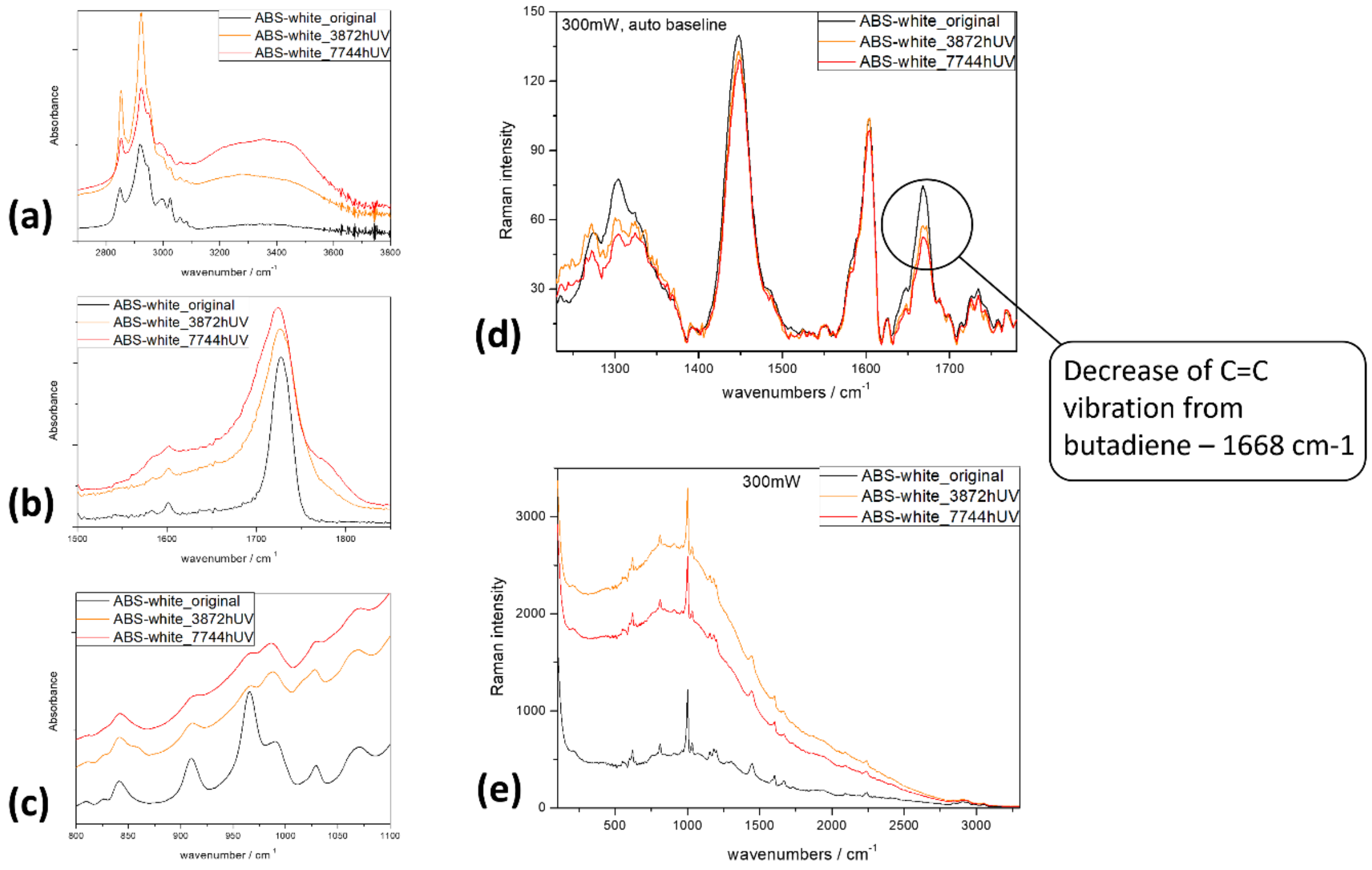

Fig. SI8. Details of as-measured vibrational spectra of ABS: (a-c) IR and (d,e) Raman spectra. Raman spectra were obtained using the 785 nm excitation laser wavelength.

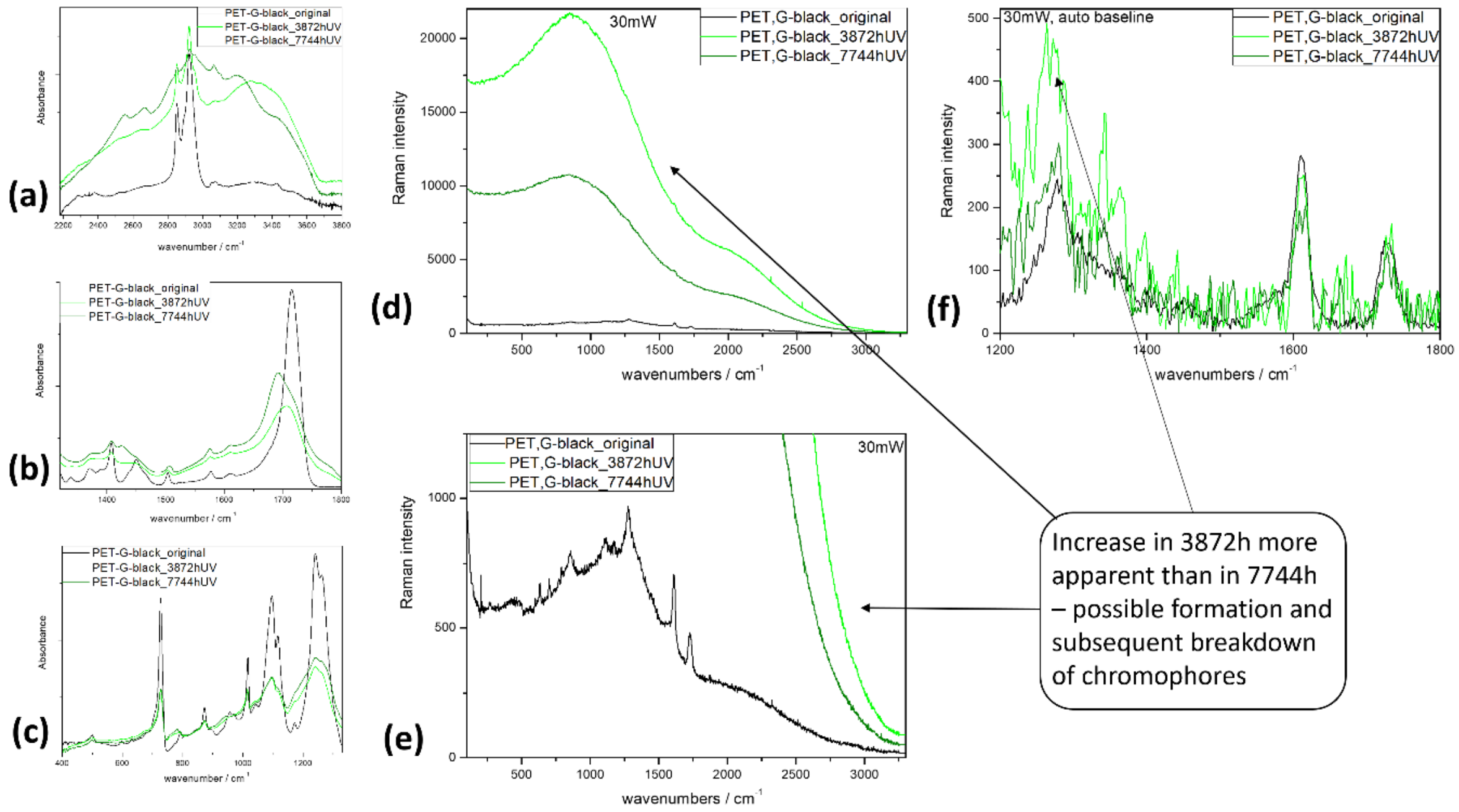

Fig. SI9. Details of as-measured vibrational spectra of PETG: (a-c) IR and (d-f) Raman spectra. Raman spectra were obtained using the 785 nm excitation laser wavelength.

### 10.4.1 Carbonyl index calculations

In order to evaluate changes in polymer structure during UV exposure, we can calculate the so-called carbonyl index (see the description below Table SI1 and/or below Table SI2). It has been done in the literature frequently, for instance in [94] for ABS and in [38] for PET.

**Carbonyl index I – in the case of ABS:**

Due to the presence of additives in our original ABS, we can perform the calculation only after subtraction of the absorbance of C=O peak present in the original ABS sample and under the assumption that the additive in our ABS does not change its own absorption in C=O region during UV-exposures). Simultaneously, we use the value of C=N absorption instead of methylene absorption because the additive can be an organic compound with its own contribution to methylene absorbance. Based on our calculations of the carbonyl index for ABS samples (see Table SI1), a slight augmentation (from 0 through 1.09 to 1.53) with the increasing time of UV irradiation is confirmed. This corresponds to the observed IR spectral changes discussed in the main text.

Table SI1: Carbonyl index calculation for ABS

| Sample name | A(C=O) | A(C=N) | carbonyl index |
|---|---|---|---|
| **ABS-original** | 0.1034 | 0.0084 | 0 |
| **ABS-3872h UV** | 0.1205 | 0.0157 | 1.09 |
| **ABS-7744h UV** | 0.1336 | 0.0197 | 1.53 |

Note: carbonyl index is calculated as: (A(C=O) of UV-treated sample minus A(C=O) of originalABS) divided by A(C=N) of a particular sample

**Carbonyl index II - in the case of PET:**

Pires et al. [38] exploited characteristic C=O vibration of esters (being characteristic for the polymeric backbone of PET only) and estimated the carbonyl index that we call carbonyl index II in this study. It is expressed in a different manner than in ABS case (expressed as carbonyl index I in this study), i.e. as the ratio of absorbance at 1720 $cm^{-1}$ vs. absorbance at 1508 $cm^{-1}$ (i.e., carbonyl of esters vs. aromatic C=C stretching which represent an unalterable band). The resulting carbonyl index in their case [38] slightly decreased for PET after the exposition to UV which means that polymeric chain scission occurred (in other words, the relative content of ester bonds decreased).

Table SI2. Carbonyl index calculation for PET-G

| Sample name | A(C=O, ester) | A(aromat. C=C) | carbonyl index |
|---|---|---|---|
| **PET-G_original** | 0.3853 | 0.0344 | 11.20 |
| **PET-G_3872h UV** | 0.1538 | 0.0390 | 3.94 |
| **PET-G_7744h UV** | 0.1827 | 0.0451 | 4.05 |

Note: carbonyl index is calculated as: (A(C=O, ester) of particular sample divided by A(aromat.C=C) of a particular sample

It should be noted on this place that there is no exact definition of the carbonyl index as already stated in [103]. Thus, it might seem puzzling that in ABS case the degradation of polymer is expressed as the carbonyl index increase, while in PET case as its decrease. In general, the carbonyl index is a measure of changes in characteristic bonds in a particular polymer, reporting thus about their degradation degree regardless the exact values being of positive or negative trend. Therefore, the carbonyl index can be expressed in different manners depending on the polymer under study. Here we dealt with ABS and PEG possessing distinctly different chemical bonds. Then, either the carbonyl index increases when C=O content is augmented during the photo-oxidation of ABS; or it can drop down (due to ester bonds cleavage) when polymeric chain scission occurs, the case of PET. Based on it, it can be recommended to define the carbonyl index for each polymer separately and discuss rather its relative changes within a given dataset. We applied this approach in our study.

Inspired by the latter approach - from ref. [38], we applied carbonyl index II calculation for our PET-G case (i.e., using the values of absorbance at 1715 $cm^{-1}$ and at 1504 $cm^{-1}$, respectively) - see Table SI2. The carbonyl index of PET-G after long-term UV exposures decreased in our case substantially in comparison to the original PET-G: from the value of approx. 11 to approx. 4. Interestingly, according to the calculated carbonyl index of PET-G, the extent of PET-G degradation induced by UV in the sample exposed for 3872h is practically same as in the sample irradiated for 7744h. The negligible difference in the carbonyl index of both UV-treated samples of PET-G may be caused by the fact that the degradation of the polymer reached a plateau after certain period of UV irradiation. Two main factors may induce such a plateau formation: (a) penetration depth of UV radiation affecting the sample properties can be reached in a certain distance from the surface and does not propagate any deeper into the sample (i.e., optical reason). (b) Measurable depth (that is constant in our ATR FT-IR setup for a given type of material) does not enable to investigate deeper layers of PET-G samples (i.e., instrumental reason).